\documentclass[showpacs,aps,prx,amsmath,superscriptaddress,twocolumn,longbibliography]{revtex4-1}

\usepackage{mathrsfs}
\usepackage{amssymb}
\usepackage{amstext}
\usepackage{txfonts}
\usepackage{mathtools}
\usepackage{color}
\usepackage[english]{babel}
\usepackage{hyperref}
\usepackage{graphicx}
\usepackage{bm}
\usepackage{float}
\usepackage{ulem}
\usepackage{dcolumn}
\usepackage{appendix}
\usepackage{comment} 

\begin{document}

\title{Topological and geometric patterns in optimal bang-bang protocols for variational quantum algorithms: Application to the $XXZ$ model on the square lattice }
\author{Matthew T. Scoggins}
\affiliation{Department of Physics and Astronomy, Western Washington University, Bellingham, Washington 98225, USA}
\author{Armin Rahmani}
\affiliation{Department of Physics and Astronomy and Advanced Materials Science and Engineering Center,
Western Washington University, Bellingham, Washington 98225, USA}
\affiliation{Kavli Institute for Theoretical Physics, University of California, Santa Barbara, California 93106, USA}
\date{\today}

\begin{abstract}

In this paper, we address the challenge of uncovering patterns in variational optimal protocols for taking the system to ground states of many-body Hamiltonians, using variational quantum algorithms. We develop highly optimized classical Monte Carlo (MC) algorithms to find the optimal protocols for transformations between the ground states of the square-lattice $XXZ$ model for finite system sizes. The MC method obtains optimal bang-bang protocols, as predicted by Pontryagin's minimum principle. We identify the minimum time needed for reaching an acceptable error for different system sizes as a function of the initial and target states and uncover correlations between the total time and the wave-function overlap. We determine a dynamical phase diagram for the optimal protocols, with different phases characterized by a topological number, namely, the number of on pulses. Bifurcation transitions as a function of initial and final states, associated with new jumps in the optimal protocols, demarcate these different phases. The number of pulses correlates with the total evolution time. In addition to identifying the topological characteristic above, i.e., the number of pulses, we introduce a correlation function to characterize bang-bang protocols' quantitative geometric similarities. We find that protocols within one phase are indeed geometrically correlated. Identifying and extrapolating patterns in these protocols may inform efficient large-scale simulations on quantum devices. 
\end{abstract}

\maketitle

\section{Introduction}

The simulation of many-body quantum states with quantum devices \cite{Feynman82} has made substantial progress. Significant efforts have focused on single-purpose quantum simulators \cite{Georgescu14}, where we physically create systems described by the model we would like to simulate. 
Adiabatic evolution is a common approach to preparing the ground state of the model Hamiltonian. If done sufficiently slowly in the absence of a vanishing spectral gap, this approach effectively prepares the desired ground state \cite{Biamonte11,Babbush2014}. However, in most cases, the target states lie across quantum phase transitions from the initial state, resulting in a vanishing gap and divergent adiabatic timescales. Furthermore, antiadiabaticity \cite{Dutta16,Ritland18} exacerbates the issue in the presence of noise.

For certain problems, nonadiabatic methods have proven promising~\cite{troyer, heim}. Despite its challenges, one promising approach for finding the ground state of many-body quantum Hamiltonians is the variational quantum algorithm (VQA). This method relies on starting from an easy-to-prepare initial state and evolving into the desired ground state of a target Hamiltonian by variationally modifying the parameters in the time-dependent Hamiltonian of the device. The idea has been explored for state preparation~\cite{Rohringer,Rosi,Rahmani13} and has showed remarkable theoretical~\cite{Peruzzo:14,Yung:14, Wecker:15, wecker, McClean:16, McClean:17,Nannicini19,Leng19,Zhou20,Nakanishi20,Kubler2020,Arrasmith20,Yao20,Wierichs20,Stokes2020} and experimental~\cite{Shen:15, Eichler:15, Omalley:15,Hempel,Otterbach,Colless,Kandala2017,Kokail2019} success, particularly in quantum chemistry simulations. It is also closely related to the quantum approximate optimization algorithm \cite{Farhi,Farhi:3, Wang20}. The scheme utilizes a hybrid quantum-classical system. Repeated physical evolutions on the quantum machine are optimized in a feedback loop to minimize the expectation value of the target Hamiltonian, thus creating the ground state of the model Hamiltonian.

There are two broad approaches to VQA, methods based on quantum circuits with parametrized gates and gate-free strategies, which may offer better coherence times~\cite{Meitei}. A version of gate-free VQA is based on quantum optimal control (QOC).  It uses a device Hamiltonian of fixed form, with the time evolution generated by varying the device's tunable parameters. The target Hamiltonian only affects the cost function, giving rise to a general-purpose simulator.
However, a large number of variational parameters and the absence of generic good initial guesses for the protocol pose challenges to this scheme. Therefore, it is crucially important to find and characterize patterns in the time dependence of tunable parameters in the Hamiltonian of the device. Possible extrapolation of these patterns to large systems may then allow efficient parametrization of the protocol to be optimized. Pontryagin's minimum principle plays a crucial role in QOC \cite{pontryagin1987, Yang17}. This minimum principle implies that, if a given set of conditions are met, the optimal path has controls that take on either their maximum or minimum value at any given time---a bang-bang protocol. The bang-bang nature makes the protocols amenable to characterization and potential extrapolation.

In this paper, focusing on the ground-state transformation of the $XXZ$ model on the square lattice, we explore optimal-protocol patterns. We search for the optimal protocols that prepare the desired target state using two different types of Monte Carlo (MC) simulations on a classical computer. The first method is direct brute-force Monte Carlo (BFMC), which does not assume bang-bang protocols, but still converges to them. Since Pontryagin's principle does not guarantee bang-bang protocols (due to the possibility of singular intervals), this inefficient algorithm is important for initial verification of the protocols' bang-bang nature. The second, bang-bang Monte Carlo (BBMC), assumes bang-bang parametrization of protocols and outperforms the BFMC in accuracy and computational efficiency. For a fixed initial and target, we find almost identical protocols for the two approaches. These optimal protocols significantly outperform the adiabatic method.

Our studies are naturally limited to small system sizes due to the computational complexity of simulating VQA on classical computers for a many-body state. Using an actual quantum device to perform the time evolution physically, we expect to access much larger systems. Nevertheless, finding the optimal protocol could still be difficult due to the complexity of the control space and the number of iterations required to reach the expectation value's global minimum. Our work aims to mitigate this issue by finding patterns in the classically obtained protocols for smaller system sizes, which we hope may inform an efficient search for optimal VQA protocols for larger system sizes. The patterns may yield an efficient parametrization upon extrapolation, helping the algorithm hone in on the optimal protocol with significantly fewer iterations.

The results of this paper are twofold. First, we develop highly efficient numerical methods for finding optimal bang-bang controls. Several improvements to the state-of-the-art algorithms are presented; these improved algorithms use adaptive moves in MC, combined discrete and continuous parametrizations, and the precompiling of unitary operators and diagonalized Hamiltonians. Second, we apply these algorithm advances to the two-dimensional $XXZ$ model. We present a full characterization of the optimal protocols for several numerically accessible system sizes and filling fractions, scanning over all initial and target ground states. In the context of our model, the exhaustive investigation allows us to raise and answer multiple new questions discussed below.

The determination of the optimal protocols for all initial and target ground states allows us to determine the total time it takes to optimally transform the ground states of a class of Hamiltonians to each other. This time serves as a practical measure of distance between all ground states, endowing the equilibrium ground states with valuable dynamical information. Furthermore, in addition to the time needed for the transformation, the associated bang-bang protocols' characteristics are of considerable interest. A salient property of bang-bang protocols is the number of square pulses in the signal. As we change the initial or target ground state, we find transitions where the number of pulses changes.

We find that the transitions mentioned above are continuous bifurcations. For example, in an interval with the control field on, an infinitesimally small interval appears, where the control field is turned off. This interval then grows continuously. We next find phase diagrams as a function of initial and target states, with different phase-diagram regions having different pulses numbers. These transitions are between distinct pulse topologies, characterized by integer numbers, so they are reminiscent of topological transitions. Furthermore, they are continuous in the sense that the duration of the new pulse emerging at a transition grows continuously from zero. We have verified that in the vicinity of the transition, the pulse durations fit power laws. 

In addition to the topological characteristic of the number of pulses, the geometric correlations between bang-bang pulses are of interest. How similar are the pulses in various regions of the space of the initial and target states? In this paper, we define a shape-shape correlation function that captures the quantitative similarity of two bang-bang protocols. Correlations and anticorrelations appear across the transitions. 

The outline of this paper is as follows. In Sec. \ref{sec:xxz}, we discuss the model and the general setup of state transformations, including the measures of distance in the optimal protocol. In Sec. \ref{sec:brute}, we discuss the brute-force MC algorithm used for an initial approximate determination of the optimal protocols. Section~\ref{sec:pontryagin} discusses Pontryagin's minimum principle and the bang-bang nature of the optimal protocols. In Sec. \ref{sec:bang}, we present our efficient algorithm for the final exact determination of the optimal bang-bang protocols. We then discuss our numerical results on the critical time needed for the optimal protocols in Sec. \ref{sec:critical_t}. In Sec. \ref{sec:topological}, we present our results on the topological phase structure of the optimal protocols and the continuous bifurcation transitions between the phases. In Sec \ref{sec:geometric}, we introduce a correlation function to capture the geometric similarities of bang-bang protocols and present results on the correlations between the protocols in one phase. Finally, we present our conclusion in Sec.~\ref{sec:conclusions}. The details of the optimized MC implementation are presented in the Appendix.

\section{model and setup}\label{sec:xxz}
\subsection{The $XXZ$ model}
In this paper, we focus our studies on the $XXZ$ model on the square lattice. Generally, in variational quantum algorithms, we can have two distinct Hamiltonian forms, the target Hamiltonian whose ground state we want to create, and the device Hamiltonian, which generates the quantum evolution of the state. However, in this paper, we focus on the case where we want to create the ground state of a Hamiltonian that has the same form as the device Hamiltonian. With this choice, the problem can be viewed as finding an optimal shortcut to the adiabatic evolution \cite{OCreview, OCrev2, OC1, OC4, Rahmani11,OC2, top3,Stefanatos2019,Stefanatos2020}, as for initial states that are also ground states for some choice of Hamiltonian parameters, adiabatic transformations are always possible in the presence of a spectral gap. Our Hamiltonian, importantly, occurs in existing systems based on superconducting qubits. We have
\begin{align*}
    H (J,K) = \sum_{\langle ij \rangle} \left[J(\sigma_i^x\sigma_j^x + \sigma_j^y\sigma_i^y) + K\sigma_i^z\sigma_j^z \right].
\end{align*}
We note that the Hamiltonian conserves $\sum_i\sigma_i^z$. The model is relevant to superconducting qubit devices~\cite{Mart}.

Due to the total $\sigma^z$ conservation, for a square lattice with $M$ sites and $C$ occupants, the Hamiltonian dimension becomes $d =$ $M \choose C$. We need a dimension of around 5000 or smaller to perform the complex optimization algorithm and find the optimal protocols. We are therefore able to explore all occupancies with a square lattice for $M \ \epsilon \ \{4,9\}$, along with some small occupancies for $M \ \epsilon \ \{16,25,36\}$ systems. We also skip the trivial cases of $C \ \epsilon \ \{0, 1\}$. Furthermore, $M-C$ occupants give rise to the same evolution as $C$ occupants due to the spin rotation symmetry. We therefore focus on occupancies $C \leqslant M/2$.

\subsection{Measures of distance for optimal control}
To prepare the ground state of the target Hamiltonian for parameters $J$ and $K$ using Monte Carlo simulations, we need to minimize a cost function. In variational quantum algorithms, the standard cost function is the expectation value of the energy. We can also define a cost function in terms of the wave function~\cite{Rezakhani09,Bao18,Friis18,}:
\begin{align*}
   {\cal C}[\psi(\tau)]_{E} &\equiv \langle \psi(\tau)|H_{\rm target}|\psi(\tau) \rangle,\\
   {\cal  C}[\psi(\tau)]_{S} &\equiv 1 - |{\langle \psi(\tau)|\psi_{\rm target} \rangle}| ^2,
\end{align*}
where $\psi(\tau)$ is  the final wave function after a total evolution\ time $\tau$. Upon successfully evolving into the target state, $ {\cal C}_{S}$ vanishes and $ {\cal C}_{E}$ attains its minimum possible value for any wave function, namely, the ground-state energy, $E_0$, of the target Hamiltonian.

Experimentally, the energy-based cost function is preferred because it is measurable even if the target ground-state wave function is \textit{a priori} unknown. We note that the ground state wave function is independent of the overall energy scale of the Hamiltonian and only depends on the ratio of the coupling constants:
\begin{align*}
    r \equiv \frac{J}{K}.
\end{align*}
Thus a unique initial and target combination is specified by two variables, $r_i$ and $r_t$.

\begin{figure}
\includegraphics[width=.49\textwidth]{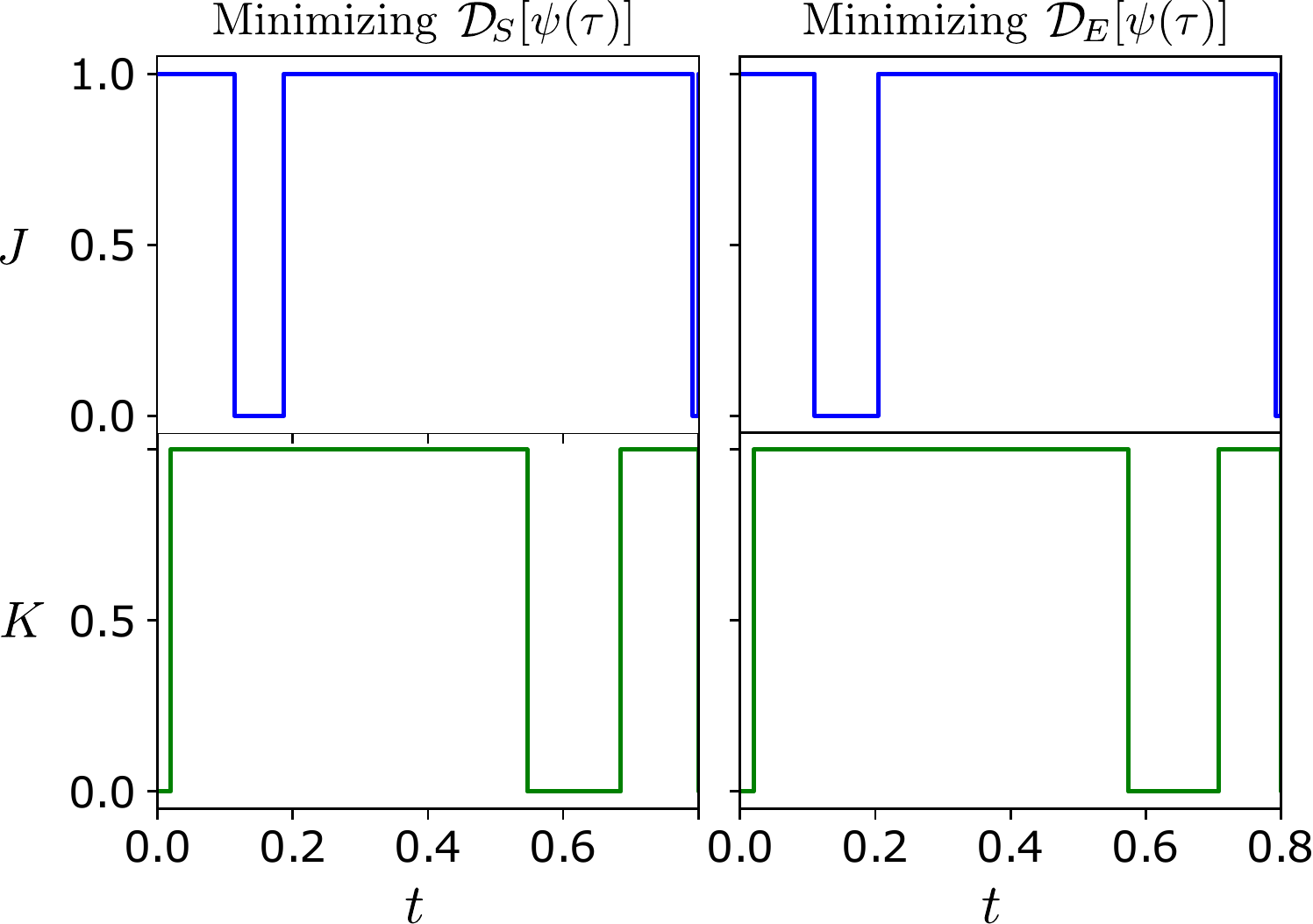}
\caption{A representative example of two different minimization schemes, achieving nearly identical protocols. $M = 9, C = 2$. 
}
\label{performance_comp1}
\end{figure}

While ${\cal  C}[\psi(\tau)]_{S}$ only depends on $r$ by construction, ${\cal  C}[\psi(\tau)]_{E}$ also depends on the energy scale of the target Hamiltonian. It is convenient to use normalized measures of distance, which are equal to 1 (0) in the initial (target) state. These can be defined in energy and state spaces as
\begin{align*}
    {\cal D}[\psi(\tau)]_{E} &\equiv \frac{\langle \psi(\tau)|H_{\rm target}|\psi(\tau) \rangle - E_0}{\langle \psi_{\rm initial}|H_{\rm target}|\psi_{\rm initial} \rangle - E_0},\\
     {\cal D}[\psi(\tau)]_{S} &\equiv \frac{1 - |{\langle \psi(\tau)|\psi_{\rm target} \rangle}| ^2}{1 - |{\langle \psi_{\rm initial}|\psi_{\rm target} \rangle}| ^2},
\end{align*} 
respectively. Clearly,  ${\cal D}[\psi(\tau)]_{E} $ is linearly related to ${\cal C}[\psi(\tau)]_{E} $, and minimizing the experimentally accessible ${\cal C}[\psi(\tau)]_{E} $ minimizes ${\cal D}[\psi(\tau)]_{E} $ . We have found that minimizing ${\cal D}[\psi(\tau)]_{E} $ and  ${\cal D}[\psi(\tau)]_{S} $ gives rise to practically identical protocols, with a representative example shown in Fig. \ref{performance_comp1}.
Hereinafter, we focus on ${\cal D}[\psi(\tau)]_{S} $ in our numerical investigations as it is customary to quantify the errors in terms of the fidelity of states, bearing in mind that a measurable energy-based cost function amenable to the variational quantum algorithms on actual quantum devices, leads to similar protocols.

We also note that for longer timescales than the time needed to reach the target state exactly, many different paths evolve into the desired target state. The optimization does not converge to unique protocols. To get the exact minimum total time, we choose to find the optimal protocols that evolve the state just short of the target state. We thus avoid convergence issues arising right at the critical time needed to reach the target state.

With the measure of distance above, we stop our Monte Carlo simulations when ${\cal D}[\psi(\tau)]_{S}  \leq 0.02$ and call the total time required to achieve the small error above, $\tau_{\rm critical}$. We can approximate the exact critical time by doing a low-order polynomial fit to the distance as a function of total time and extrapolate the time where ${\cal D}[\psi(\tau)]_{S}  = 0$. The extrapolation of these protocols yields very similar protocols, characterized by minor, unimportant modifications.

\section{Brute-Force Monte Carlo Method}
\label{sec:brute}
To find the optimal protocol and shortcut the adiabatic method, we first use a brute-force Monte Carlo (BFMC), previously used in several publications~\cite{Rahmani11,Yang17,Bao18,Jones}. In this approach, we discretize time into identical fixed intervals and allow the protocols to take on any value within the bounds of our parameters, in this case [0, 1]. With $N$ intervals, the final state is
\begin{align}
    |\psi(\tau)\rangle = \prod_j^N\ e^{-i\frac{\tau}{N}H(J_j,K_j)}|\psi_{\rm initial}\rangle.
\end{align}

{\color{black}
The specific algorithm used is simulated annealing, where implementation requires a random initial protocol $\{J_i\}_0, \{K_i\}_0$ and a pseudotemperature $T$ that decreases with the progression of the algorithm. This pseudotemperature $T$ controls the probability that nonoptimal changes are accepted, which prevents the algorithm from being stuck in local minima. We pick an initial pseudotemperature $T_0$ to have an initial acceptance rate of around $85\%$ for changes in the protocol that increase the cost $\cal C$, which can be calculated by numerically sampling random changes in the protocol. We also initially run the simulations for a smaller total time than the evolution time and slowly increase $\tau$ to the desired value as the simulations progress. We then follow this simulated-annealing procedure:

\begin{enumerate}
    \itemsep-0.2em 
    \item Change the value of the protocol at a random time step by some small amount randomly selected from the interval $[0,\frac{T}{T_0}]$. 
    \item Repeat the evolution, and measure the new cost $\cal C_{\rm new}$
    \item If this value is smaller than the previous cost, keep the change. Otherwise, keep the change with probability $\exp\left[-\frac{1}{T}(\cal C_{\rm new}-\cal C_{\rm old})\right]$.
    \item Repeat steps 1-3 for $N_{\rm sweeps}$ sweeps, then reduce $T$ (we decreased $T$ by $5\%$, i.e., $T\to 0.95 T$).
    \item Repeat steps 1-4 $N_{\rm decay}$ times, calculating $N_{\rm decay}$ to allow $T$ to get close to 0. Set $T=0$ and run $N_{\rm frozen}$ more times, then increase $\tau$.
    \item Repeat steps 1-5 until $\mathcal{D}[\psi(\tau)] \leq \epsilon$ for some allowable error $\epsilon$. In our case, $\epsilon = 0.02$. 
\end{enumerate}
}
This algorithm is inefficient as it does not utilize the bang-bang nature of the optimal protocols. However, due to the possibility of singular intervals, Pontryagin's minimum principle does not guarantee bang-bang protocols. This brute-force search is necessary for verifying that the protocols are indeed bang-bang. The piecewise-constant parametrization is more suitable for finding bang-bang protocols than other parametrizations such as a truncated Fourier series.

\begin{figure}
\includegraphics[width=0.5\textwidth]{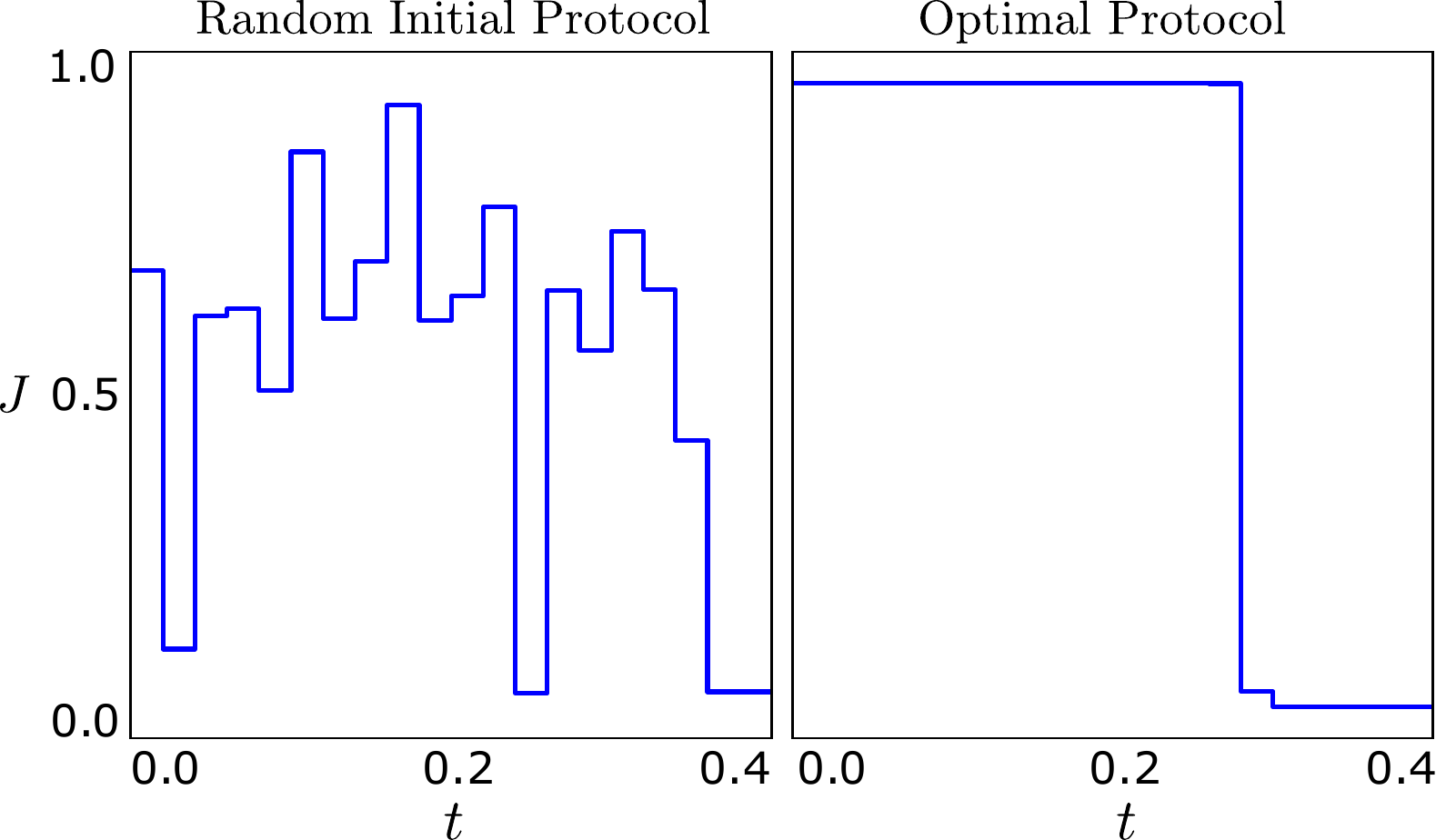}
\caption{A random initial protocol vs the optimal post-BFMC protocol for parameter $J$.}
\label{init_vs_opt}
\end{figure}

The iteration limits $N_{\rm sweeps}, N_{\rm decay}, N_{\rm frozen}$ should be chosen to get sufficiently close to the optimal protocol for each $\tau$. 
To have confidence that we are reaching the optimal protocol for each $\tau$, we repeat the process for multiple seeds that create different initial protocols and changes throughout the process but converge on the same protocol. 
This BFMC process is also repeated for a different number of intervals, $N$ until an increase in $N$ creates a negligible difference in convergence. For our case, $N = 20$ was sufficient. We find that the protocols indeed collapse into bang-bang protocols, approaching either the maximum or the minimum value (1 or 0) shown in Fig. \ref{init_vs_opt}. 

\begin{figure}
\includegraphics[width=.49\textwidth]{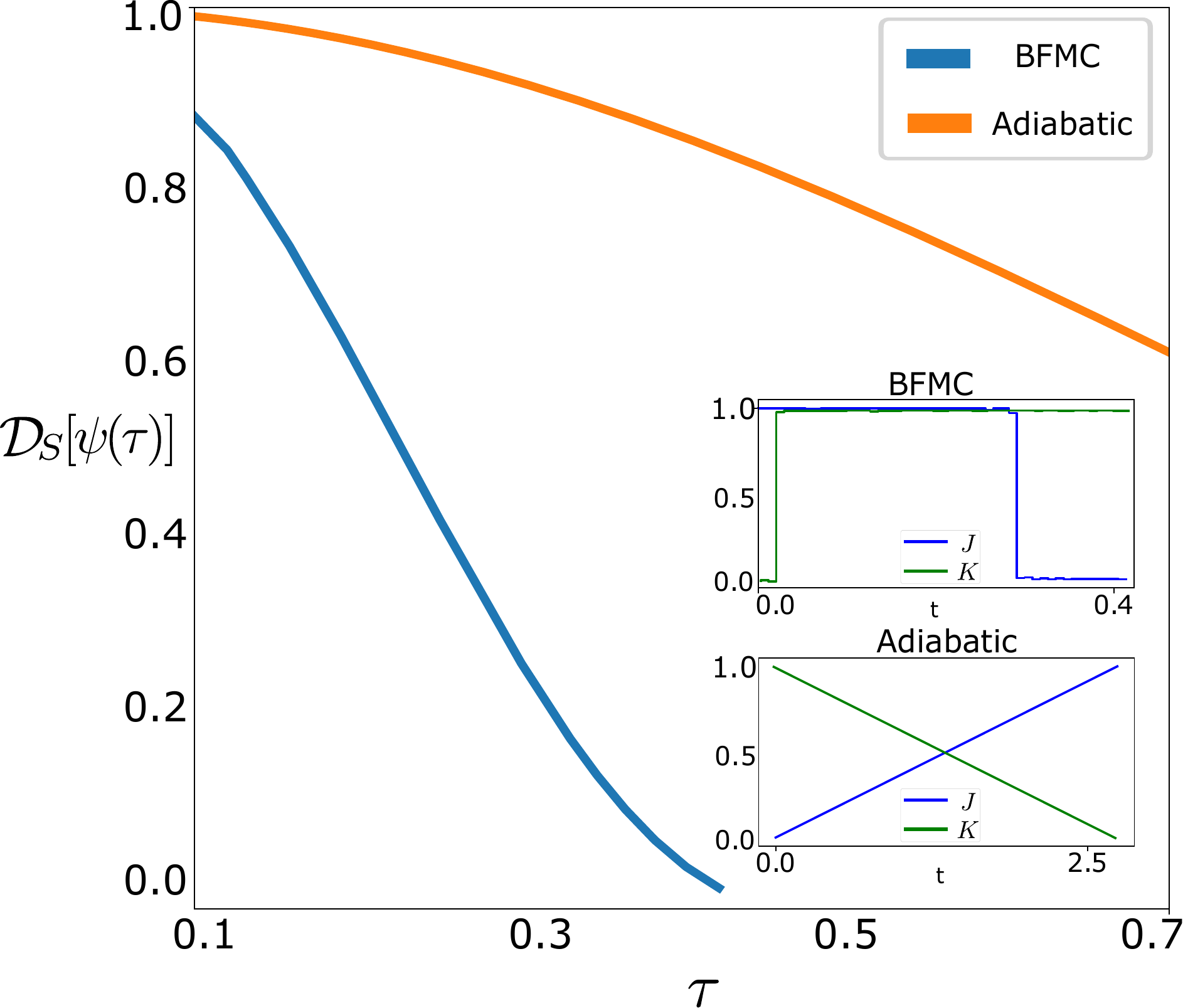}
\caption{An example of the distance vs  $\tau$ for the two methods with $M=2,\ C=2,\ r_i = 0.11,\ r_t = 9$. The BFMC achieves the ground state in a much shorter time. The optimal protocol for $D_S[\psi(\tau)] = 0$ is shown in the insets.
}
\label{performance_comp2}
\end{figure}

It is illuminating to compare the performance of these optimal protocols with the adiabatic method. Evolving from an initial to a target state can be carried out adiabatically by smoothly changing the controls into the controls corresponding to the target state. If done sufficiently slowly in the absence of a vanishing spectral gap, this approach prepares the desired ground state. 
We choose a linear time dependence for the Hamiltonian parameters. The results are shown in Fig.~\ref{performance_comp2} and show a substantial difference in the absolute error in the vicinity of the critical time for optimal evolution.
\section{Pontryagin's Minimum Principle}
\label{sec:pontryagin}
{\color{black}
Pontryagin's minimum principle is a theorem in applied mathematics that predicts generically bang-bang protocols for linear control functions. Here, we briefly review the formalism. Consider a set of dynamical variables $\bm x$, which evolve with a first-order differential equation $\dot {\bm x}={\bm f}({\bm x}, \boldsymbol{  g})$ that contains certain time-dependent parameters $\boldsymbol{  g}(t)$. Given the initial values of the dynamical variables ${\bm x}(0)$, the differential equation determines their final values for each set of time-dependent control parameters. Suppose we want the optimal controls $\boldsymbol{  g}^*(t)$ that minimize a function ${\cal F}[{\bm x}(\tau)]$ of the dynamical variables at the final time $\tau$. Pontryagin's minimum principle states that 
\begin{equation}\label{eq:pont}
{\cal H}({\bm x}^*,{\bm p}^*,\boldsymbol{  g}^*)=\min_{\boldsymbol{  g}}{\cal H}({\bm x}^*,{\bm p}^*,\boldsymbol{  g})
\end{equation}
for any time $0<t<\tau$, where we have defined conjugate momenta $\bm p$ that evolve as $\dot {\bm p}=-\partial_{{\bm x}}{\cal H}$ with boundary conditions ${\bm p}(\tau)=\partial_{{\bm x}}{\cal F}[{\bm x}(\tau)]$ and the optimal-control Hamiltonian ${\cal H}({\bm x},{\bm p},\boldsymbol{  g})\equiv {\bm f}({\bm x}, \boldsymbol{  g})\cdot{\bm p}$.  In Eq. ~\eqref{eq:pont}, ${\bm x}^*$ and ${\bm p}^*$ represent the solutions for the dynamical variable and their conjugate momenta, respectively, corresponding to the optimal controls $\boldsymbol{  g}^*(t)$. If the equations of motion are linear in $\boldsymbol{  g}(t)$, then the optimal-control Hamiltonian will be a linear function of $\boldsymbol{  g}(t)$, and Eq.~\eqref{eq:pont} indicates that $\boldsymbol{  g}^*(t)$ takes its minimum or maximum allowed value at every point in time, leading to bang-bang protocols.

Now consider a general quantum state evolving with the Schr\"odinger equation $\partial_t |\psi(t)\rangle=-i H(t)|\psi(t)\rangle$. The Hamiltonian contains some tunable coupling constants $g_\alpha(t)$, which we can change as a function of time:
\begin{equation}
    H(t)=\sum_\alpha g_\alpha(t) O_\alpha,
\end{equation}
where $O_\alpha$ are some Hermitian operators. We can tune each of the coupling constants in some range \begin{equation}
g^{\min}_\alpha<g_\alpha(t)<g^{\max}_\alpha.\end{equation}

Apart from the constrained range above, we assume that we can impart an arbitrary time dependence to the coupling constant, to transform the initial state $|\psi(0)\rangle$ into the target state $|\psi_{\rm target}\rangle$. This can be achieved by fixing the total time of the evolution, $\tau$, and minimizing the cost function ${\cal C}(|\psi(\tau)\rangle)=1-|\langle\psi(\tau)|\psi_{\rm target}\rangle|^2$.

To apply Pontryagin's minimum principle, we consider all the amplitudes needed to specify the wave function $|\psi(t)\rangle$ in an orthonormal basis as our dynamical variables $\bm x$. For the  conjugate momenta $\bm p$, we define a conjugate state $|\Pi(t)\rangle$ that evolves with the same Schr\"odinger equation $\partial_t |\Pi(t)\rangle=-i H(t)|\Pi(t)\rangle$. Unlike the quantum state whose boundary condition is known at the initial time, the conjugate states have known boundary conditions at the final time
\begin{equation}
    |\pi(\tau)\rangle=\partial_{\psi}{\cal C}(|\psi\rangle)\big|_{t=\tau}, 
\end{equation}
where ${\cal C}(|\psi(\tau)\rangle)$ plays the role of ${\cal F}[{\bm x}(\tau)]$ of the general formalism. The above derivative should be interpreted in terms of the real and imaginary parts of the components of $\psi$. 
For our particular fidelity-based cost function, we have
\begin{equation}
    |\Pi(\tau)\rangle=-2|\psi_{\rm target}\rangle\langle\psi_{\rm target}|\psi(\tau)\rangle.
\end{equation}}
The state and its conjugate determine whether the controls take their minimum or the maximum allowed values according to~\cite{Jones}
\begin{equation}
   g_\alpha(t)=\left\{\begin{array}{c}
g_\alpha^{\max}, \quad {\rm Im}[\langle \Pi(t)|O_\alpha|\psi(t)\rangle]<0 \\ 
g_\alpha^{\min}, \quad {\rm Im}[\langle \Pi(t)|O_\alpha|\psi(t)\rangle]>0.
\end{array} \right.
\end{equation}

\begin{figure}[h!]
\includegraphics[width=0.5\textwidth]{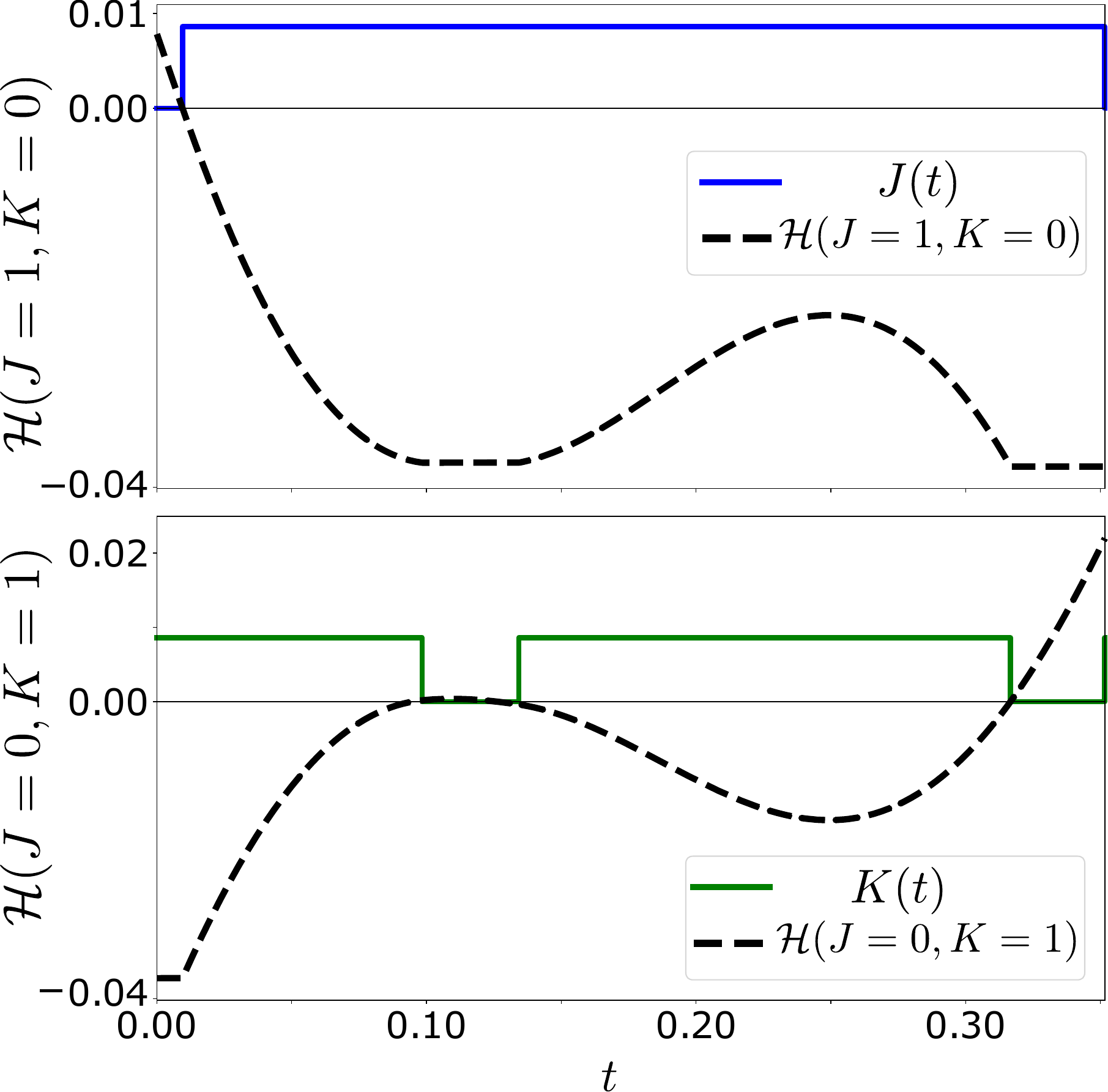}
\caption{An example of $\mathcal{H}$ for $M = 9,\ C = 2,\ r_i = 4.752,\ r_t = 0.582$.}
\label{fancy_h}
\end{figure}

In our case, the Hamiltonian has two tunable coupling constants $J$ and $K$,
and we can write $O_J=H(J=1, K=0)$ and $O_K=H(J=0, K=1)$. An example is shown in Fig.~\ref{fancy_h}. The flat pieces in the figure are a consequence of the evolution generated by a Hamiltonian $H=O_\alpha$ in these intervals, which gives $e^{iHt}O_\alpha e^{-iHt}=O_\alpha$.

\section{Bang-bang Monte Carlo techniques}
\label{sec:bang}
With the bang-bang nature of the protocols confirmed, we take advantage of this form and create more efficient Monte Carlo processes, allowing exploration of larger system sizes that were previously limited due to infeasible computing times. Computationally, the unitary operator generation is by far the most demanding part of the simulation, taking ${O}(d^3)$ where $d = $ $M\choose C$ is the dimension of the Hamiltonian. Of course, this step is precisely what the quantum device will perform by physical evolution and measurement instead of calculating the solution to the Schr\"odinger equation on a classical computer. In our investigation on a classical computer, however, we need to reduce the number of times we generate the unitary operator to make the simulations more efficient. We run a two-step bang-bang Monte Carlo (BBMC) algorithm. First, we apply the discrete-bang Monte Carlo (DBMC) algorithm, which is similar to the BFMC, but the protocols are restricted to the maximum and minimum within our parameter range, 1 and 0. After that, we apply the continuous-bang Monte Carlo (CBMC) algorithm, which changes the simulation parameter to when transitions occur, avoiding restricting the jumps to discrete intervals. 

\subsection{Part 1: Discrete-bang Monte Carlo}
{\color{black}The DBMC avoids the expensive unitary operator generation at each step in the evolution by precompiling the unitary operators once for each time step $\frac{\tau}{N}$ for $N$ total intervals. The protocol is parametrized as a piecewise constant protocol, where the control parameter for each interval is set either at either the minimum or the maximum allowed value instead of searching over all intermediate values, utilizing the result of Pontryagin's principle. The computations then resemble Monte Carlo simulations of an Ising-type system. For a single timestep, we are only required to generate three $U_{JK}$, operators $U_{11},U_{10},$ and $ U_{01}$, where the subscript indicates the constant values of $J$ and $K$ over a time $\tau/N$. For example, $U_{10}$ corresponds to an interval where $J$ takes its maximum value and $K$ is turned off. }We note that $U_{00} = I$ and should not appear in any optimal protocols since its only effect is wasting time without changing the state. Then, each step in the evolution is reduced to $O(d^2)$ matrix-vector multiplication. We also take advantage of adaptive step sizes for a given $\tau$, allowing us to start with a coarse protocol, i.e., small $N$, and iteratively double the number of intervals for a fixed $\tau$. For small $N$, optimization is computationally inexpensive but typically far from the true optimal protocol. For large $N$, convergence requires many sweeps if starting from a random initial protocol. This adaptive method, where the initial protocol for larger step sizes is generated by the optimized protocol for the previous step size, significantly reduces the total number of sweeps required for convergence.

\subsection{Part 2: Continuous-bang Monte Carlo}
{\color{black}In this approach, a certain number of jumps are assumed and the corresponding times for these jumps are treated as the variational parameters of the protocol. This number is typically very small (less than 5) so we are left with a simulation with very few variational parameters. Of course, the results of the DBMC provide a good estimate for the number of jumps and their approximate time. With a continuous parameter, namely, the time of each pulse, treated as a variational parameter, we cannot precompile the operators and must generate the unitary operators at each step in the evolution. However, the CBMC shortcuts this generation by prediagonalizing $H_{JK}$ for the three possible combinations of $J$ and $K$, saving the eigenvectors and eigenvalues $V_{JK}$ and $D_{JK}$, and expressing the unitary operator as $ U(\triangle t) = V_{JK}e^{-i\triangle tD_{JK}}V_{JK}^\dagger$ for timestep $\triangle t$.} Then, the only time-dependent component which must be generated at each step in the evolution is $e^{-i\triangle t D_{JK}}$, which takes $\mathcal{O}(d)$ operations. We then evolve the state according to $|\psi(t + \triangle t)\rangle = V_{JK}e^{-i\triangle tD_{JK}}V_{JK}^\dagger |\psi(t)\rangle$, where we avoid matrix-matrix multiplication by doing three matrix-vector multiplications. This approach reduces the evolution down to ${O}(d^2)$ operations. This approach allows for true optimal convergence due to avoiding the interval restriction. It is also quite efficient, particularly when combined with the first discrete step that effectively determines the number and approximate jumps' location.

{\color{black}This technique outperforms the BFMC in optimal-protocol accuracy and computational efficiency. The performance gains are substantial. For small systems accessible to BFMC, the running times are improved by around three to four orders of magnitude, reducing the total computation time for all initial and target states from weeks to minutes. For larger systems, the computations become infeasible with the BFMC algorithm. Thus our BBMC method gives access to system sizes with Hilbert spaces of dimension up to around 5000 with our computing power. We compare the protocols found from this simulation with the BFMC in Fig. \ref{mc_comps}; they are nearly identical.}
 We discuss several more algorithm optimizations in the Appendix.

As our ultimate goal is to search for patterns in the optimal protocols across system sizes, different protocols must achieve the same measurement of distance $\cal D$. Scaling $\tau$ makes it unlikely that two different initial-target combinations will have the same $\cal D$. So, after achieving ${\cal D} \leq 0.02$, we implement a binary search in $\tau$ which hones in on the total time required to achieve the optimal protocol ${\cal D} = 0.02$.

\section{Properties of the critical total time}
\label{sec:critical_t}
We first present our numerical results for the critical total time $\tau_{\rm critical}$ for reaching the target. The data are presented in a color plot with the horizontal (vertical) axis representing the initial (target) state in terms of the parameters $\ln(r_i)$ and $\ln(r_t)$. We explore a wide range of parameters with either $J$ or $K$ dominating.

For a fixed initial and target state, a perfect optimal evolution with ${\cal D}=0$ has an evolution determined by $|\psi_{r_t}\rangle = U|\psi_{r_i}\rangle$, which means the optimal evolution from $|\psi_{r_t}\rangle$ into $|\psi_{r_i}\rangle$ can be done with the same protocol running backwards in time. Therefore the total evolution time and other quantities calculated in this paper (including the number of pulses and characteristic pulse time) are symmetric about the diagonal in the $(r_i,r_t)$ space. Although we use ${\cal D}=0.02$, and despite possible numerical artifacts and inaccuracies, we indeed observe this symmetry, confirming that we are finding very similar optimal protocols to those that prepare the target state exactly.

\begin{figure}
\includegraphics[width= 0.495\textwidth,height = 0.45\textheight]{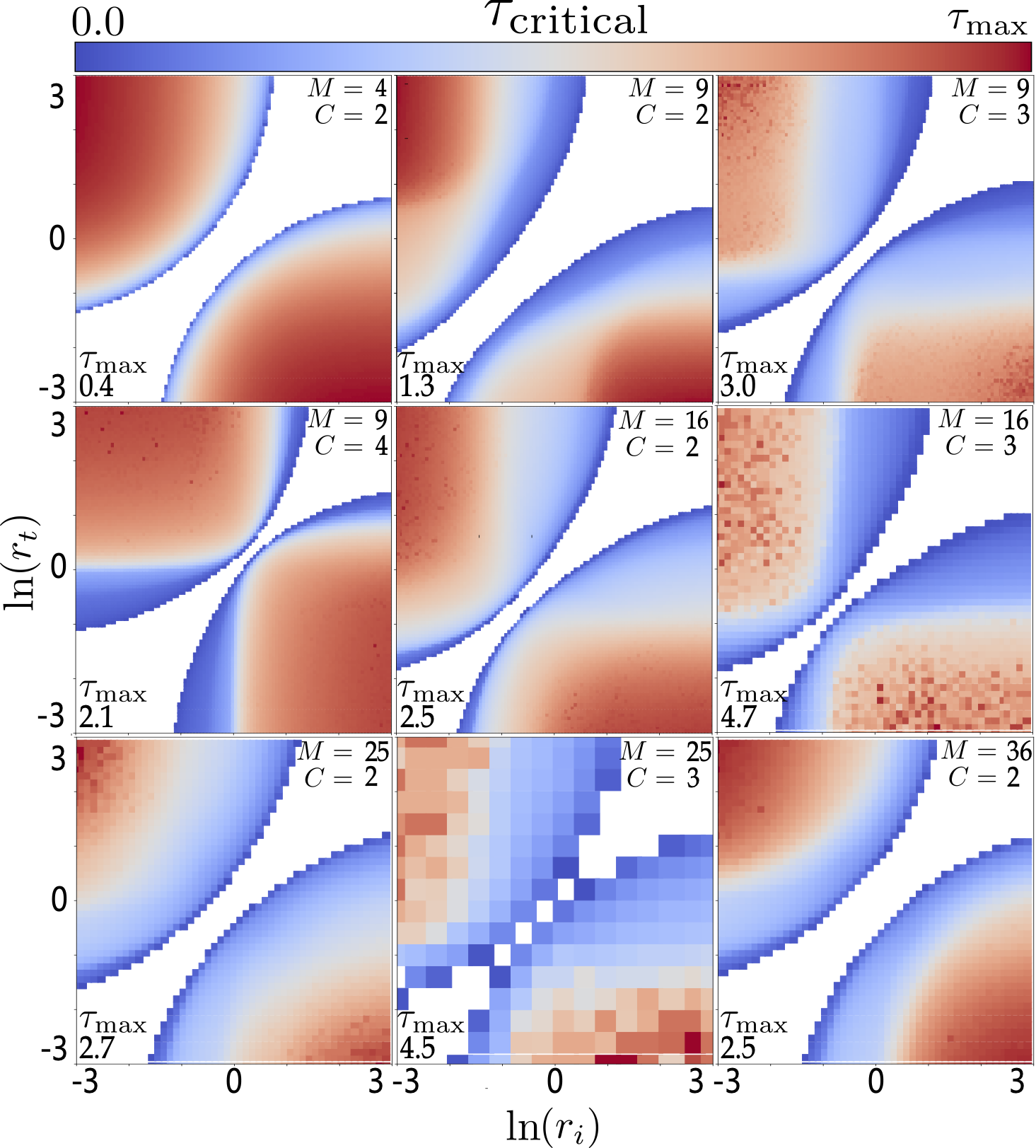}
\caption{$\tau_{\text{critical}}$ for all nine system sizes explored. As the dimension $d =$ $M\choose C$ increases, we decrease the resolution due to computational complexity. White space indicates no data due to the initial and target states being nearly identical}
\label{tau_plots1}
\end{figure}

\begin{figure}
\includegraphics[width= 0.495\textwidth,height = 0.45\textheight]{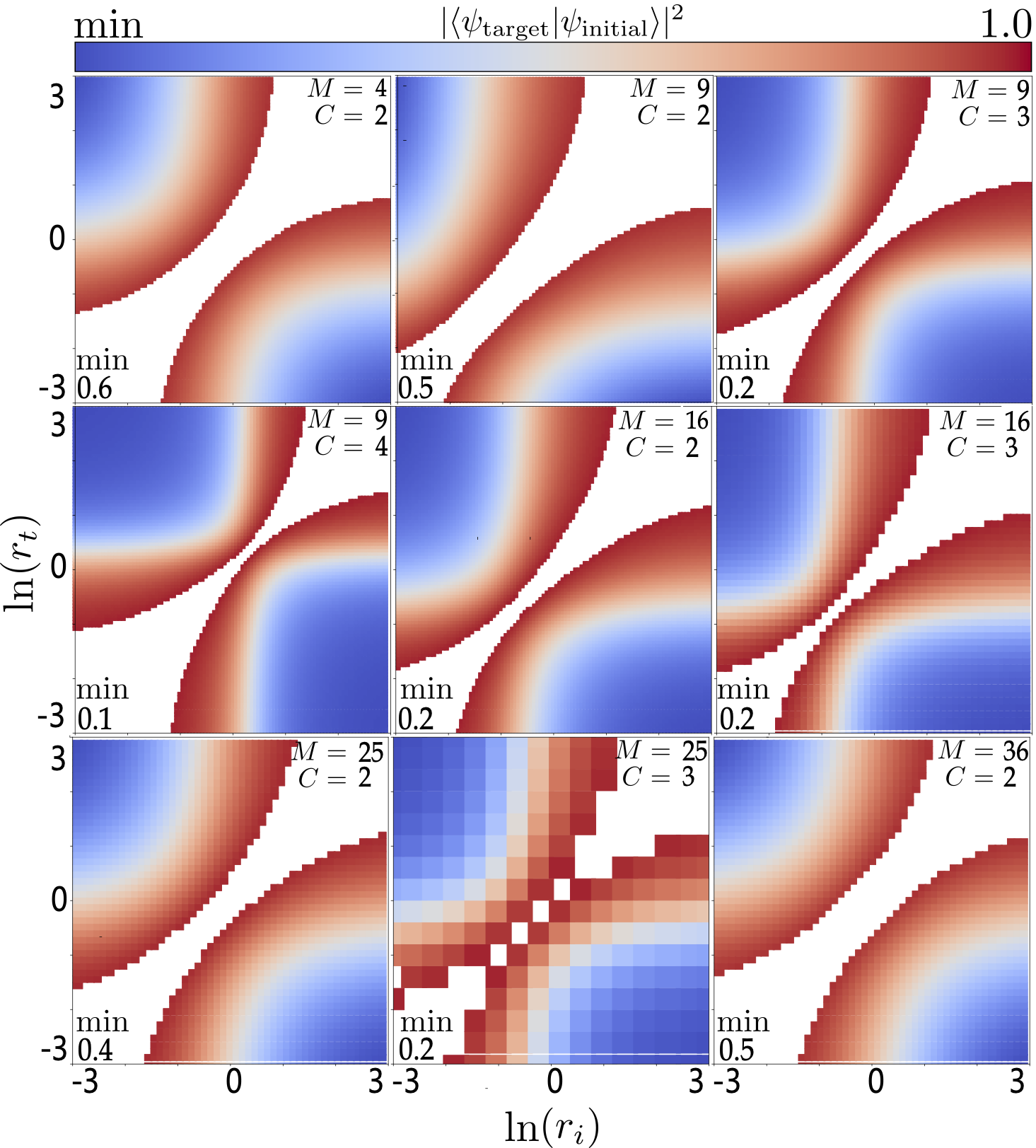}
\caption{A decrease in $|\langle \psi_{\text{\rm target}}| \psi_{\text{initial}}\rangle|^2$ increases $\tau_{\text{\rm critical}}$ which in turn increases the number of pulses in the optimal protocol shown in Figure \ref{jumps}. As the dimension $d =$ $M\choose C$ increases, we decrease the resolution due to computational complexity. White space indicates no data due to the initial and target states being nearly identical.}
\label{target_init_comp}
\end{figure}

\begin{figure}[h!]
\includegraphics[scale=0.6]{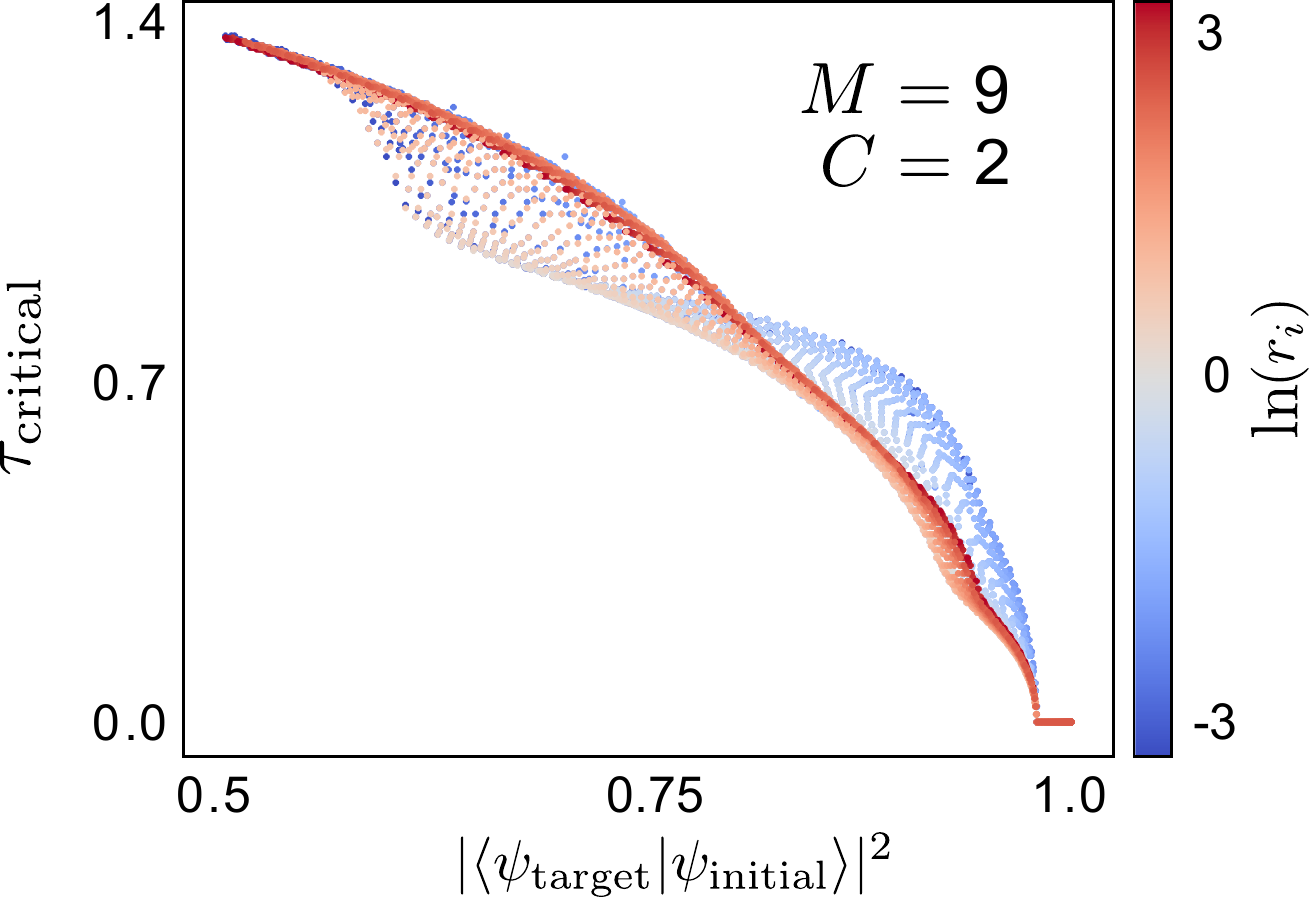}

\vspace{0.2cm}

\includegraphics[scale=0.6]{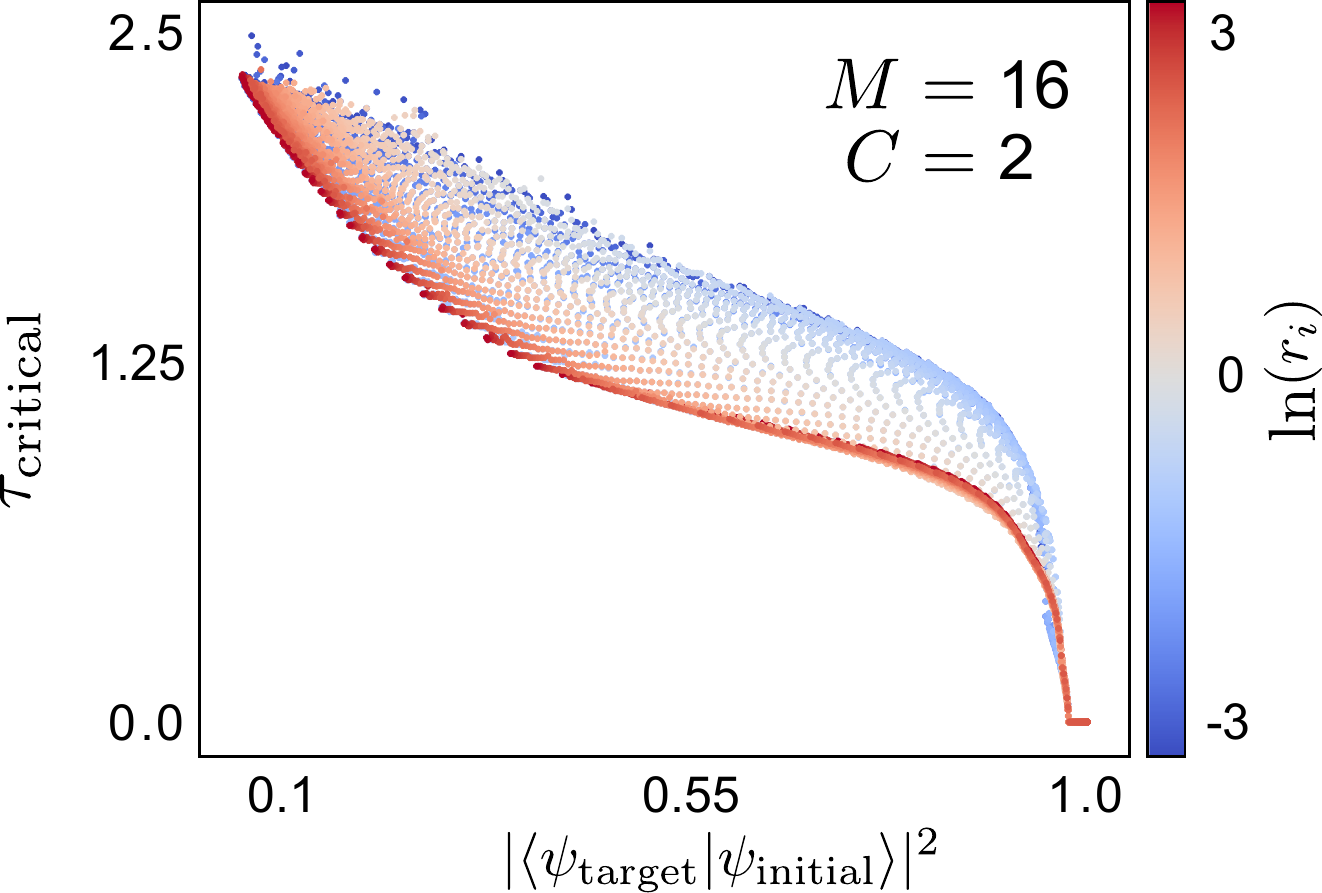}
\caption{$\tau_{\rm critical}$ as a function of initial-target overlap.}
\label{overlap_vs_tau}
\end{figure}

Patterns emerge in $\tau_{\rm critical}$ across all system sizes, as shown in Fig. \ref{tau_plots1}. As $(r_i, r_t)$ gets further away from the diagonal, $\tau_{\text{critical}}$ increases, as expected. This increase correlates with a decrease in $|\langle \psi_{\rm target}|\psi_{\rm initial}\rangle|^2$, and this overlap is shown in Fig. \ref{target_init_comp}. Intuitively, increasing the distance between the initial and target states should increase the total time. Figure \ref{overlap_vs_tau} directly shows the relationship between $\tau_{\rm critical}$ and $|\langle \psi_{\rm target}|\psi_{\rm initial}\rangle|^2$ for two different system sizes. For a fixed $r_i$, there is a clear correlation between the two.

\begin{figure}
\includegraphics[width=.49\textwidth, height=0.2\textheight]{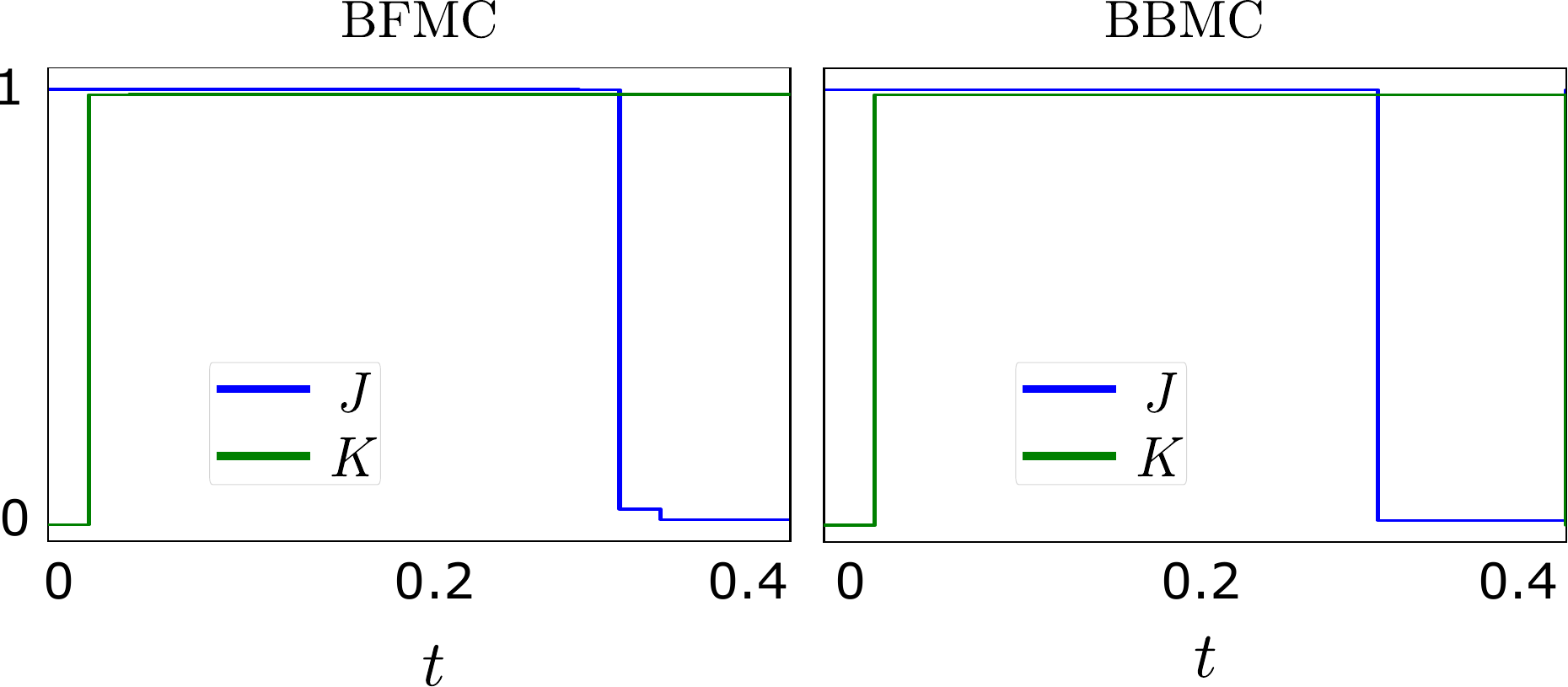}
\caption{Examples of the final optimal protocols for the two MC methods. Assuming bang-bang protocols achieves the same shape, but performs slightly better in $\mathcal{D}[\psi(\tau)]$.}
\label{mc_comps}
\end{figure}

\begin{figure}[h!]
\includegraphics[width=0.45\textwidth,height = 0.3\textheight]{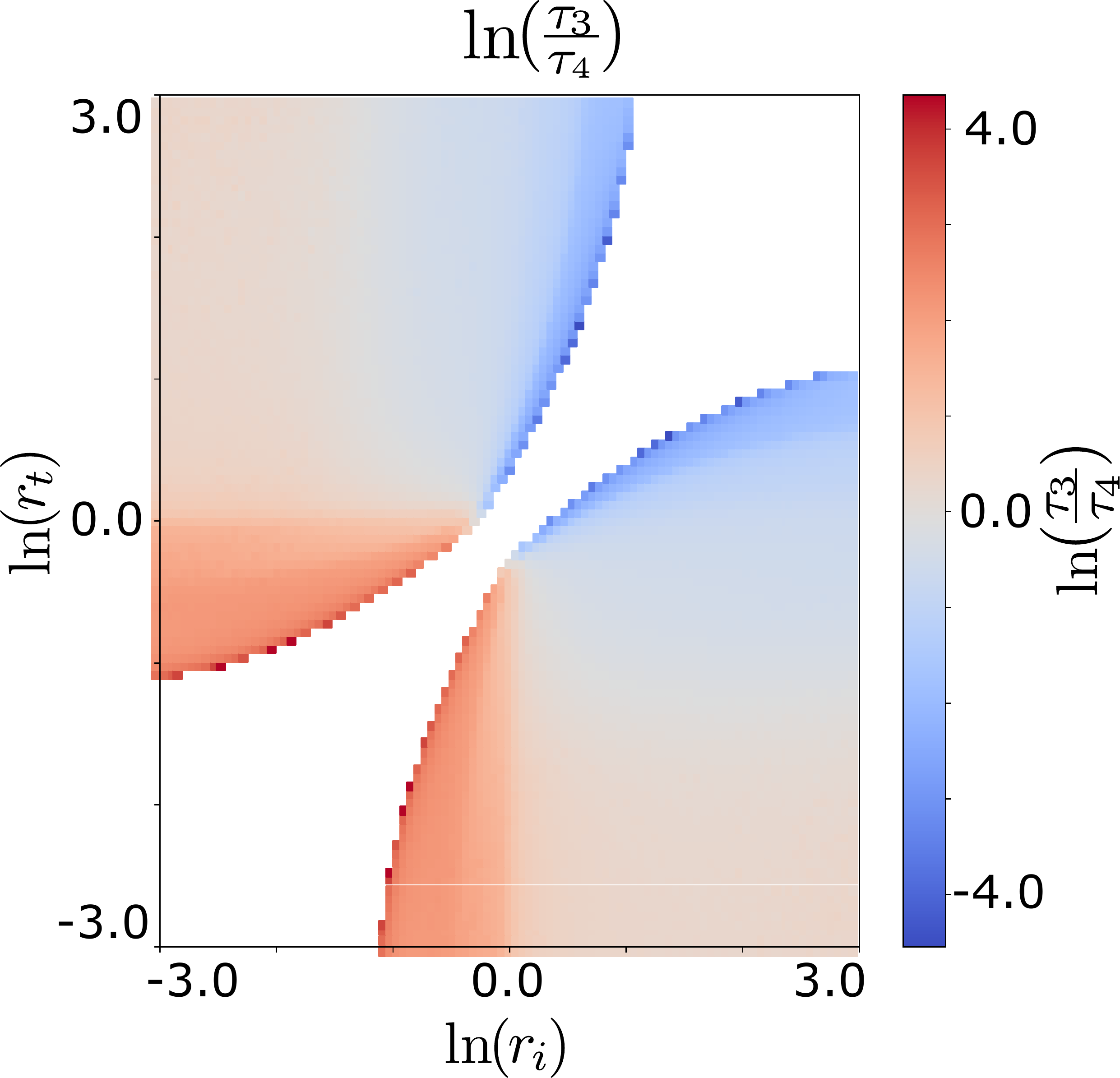}
\caption{The log of the ratios of $\tau$ for $M = 9$ between $C = 3$ and $C=4$ occupants across all combinations of $r_i, r_t$. For $r_i \leq 1, r_t \leq 1$, most optimal protocols for three occupants had total times which were greater than those for four occupants. $\overline{\tau_3} = 1.11 \pm 0.92$,  $\overline{\tau_4} = 0.99 \pm 0.78$} 
\label{tau_ratios_34}
\end{figure}

An important finding of these numerical studies concerning the promise of applying them to actual hybrid classical-quantum devices for VQA involves the dependence of the critical time on the Hilbert space dimension. 
Although systems with a larger Hilbert space lead to an increase in classical computing time, we sometimes find a shorter $\tau_{\rm critical}$ in a larger Hilbert space. As shown in Fig. \ref{tau_ratios_34}, for $\ln(r_i) < 0, \ln(r_t) < 0$ we see that $\tau_3 > \tau_4$, where $\tau_C$ is for  $M=9$ with $C$ occupants. $\overline{\tau_3} = 1.11\pm 0.92$ and $\overline{\tau_4} = 0.99 \pm 0.78$. {\color{black}The correlation of the wavefunction overlap with the total time plays an important role here. Although, when $\tau_3>\tau_4$, the $C=4$ system does not always have a larger overlap between the initial and the target states than the $C=3$ system, in most of the darker red region where $\tau_3\gg\tau_4$, there is indeed a larger overlap between the initial and target states for the $C=4$ system.} This result implies that the complexity of the VQA does not necessarily increase as the fully classical counterpart becomes exponentially more expensive, suggesting a path to quantum supremacy for the determination of many-body ground states using optimal control.

\begin{figure*}
\includegraphics[width= 0.495\textwidth,height = 0.44\textheight]{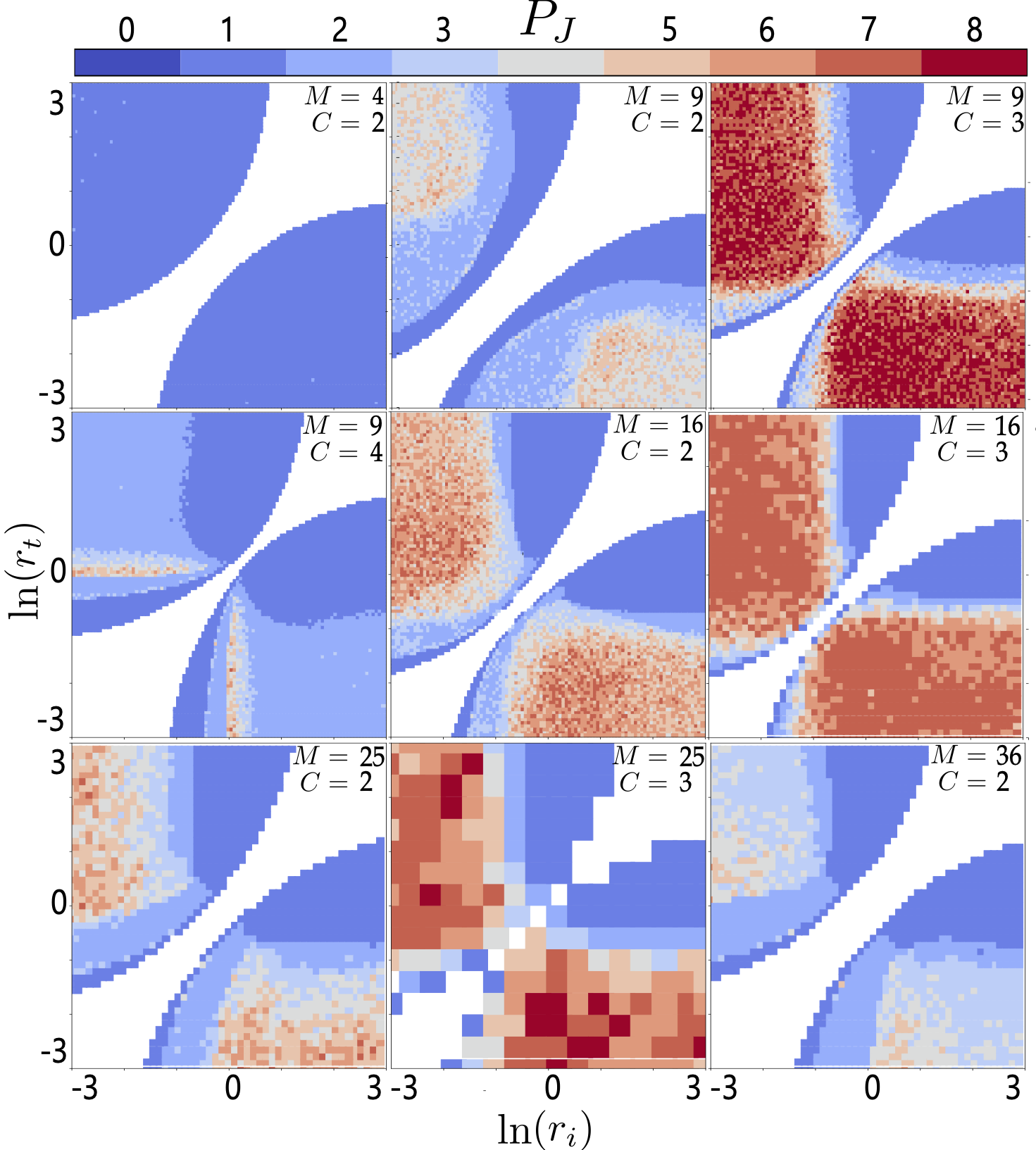}
\includegraphics[width= 0.495\textwidth,height = 0.44\textheight]{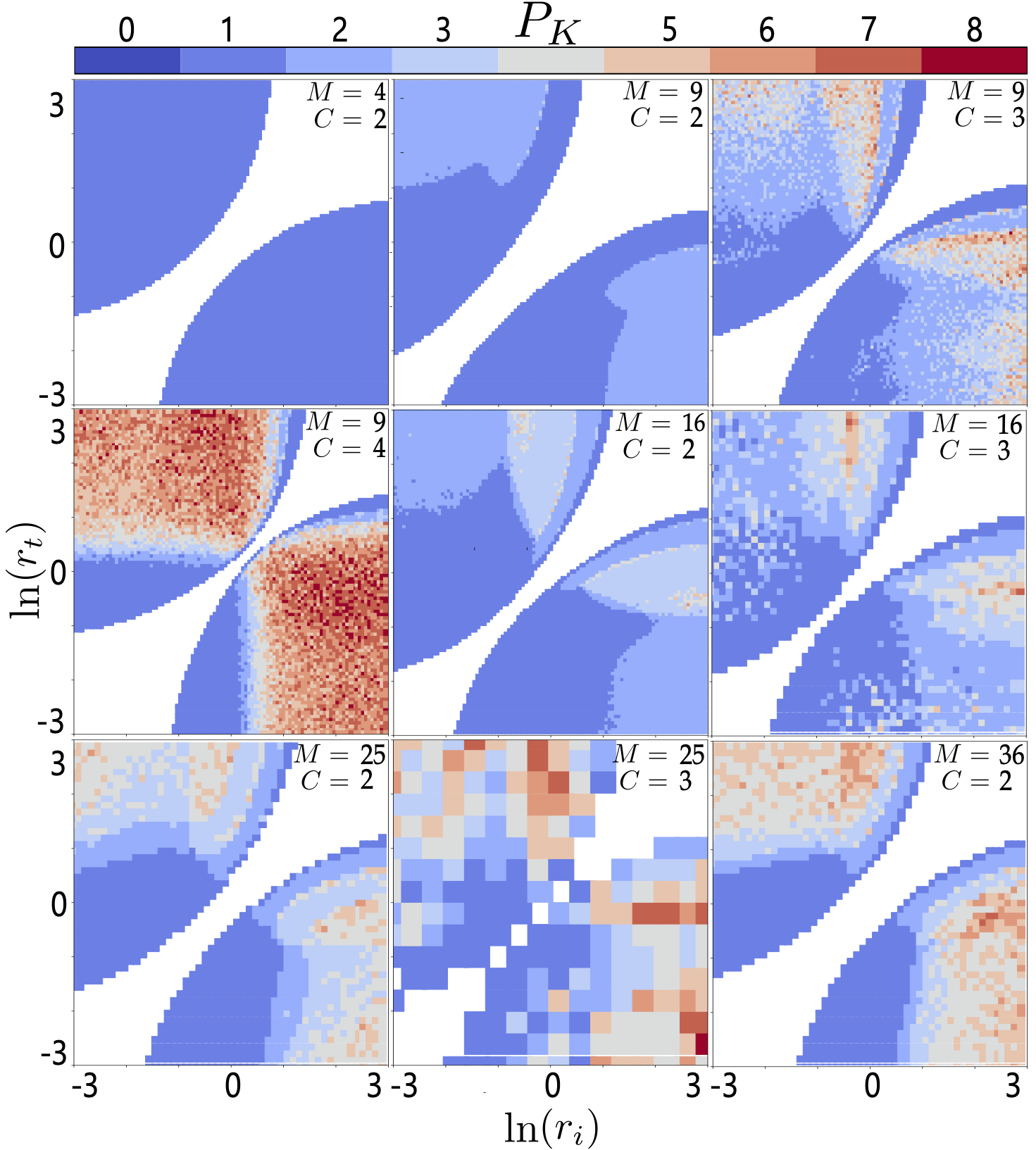}
\caption{The number of pulses for all nine system sizes explored. As the dimension $d =$ $M\choose C$ increases, we decrease the resolution due to computational complexity. White space indicates no data due to the initial and target states being nearly identical.}
\label{jumps}
\end{figure*}

\begin{figure*}
\includegraphics[width= 0.495\textwidth,height = 0.44\textheight]{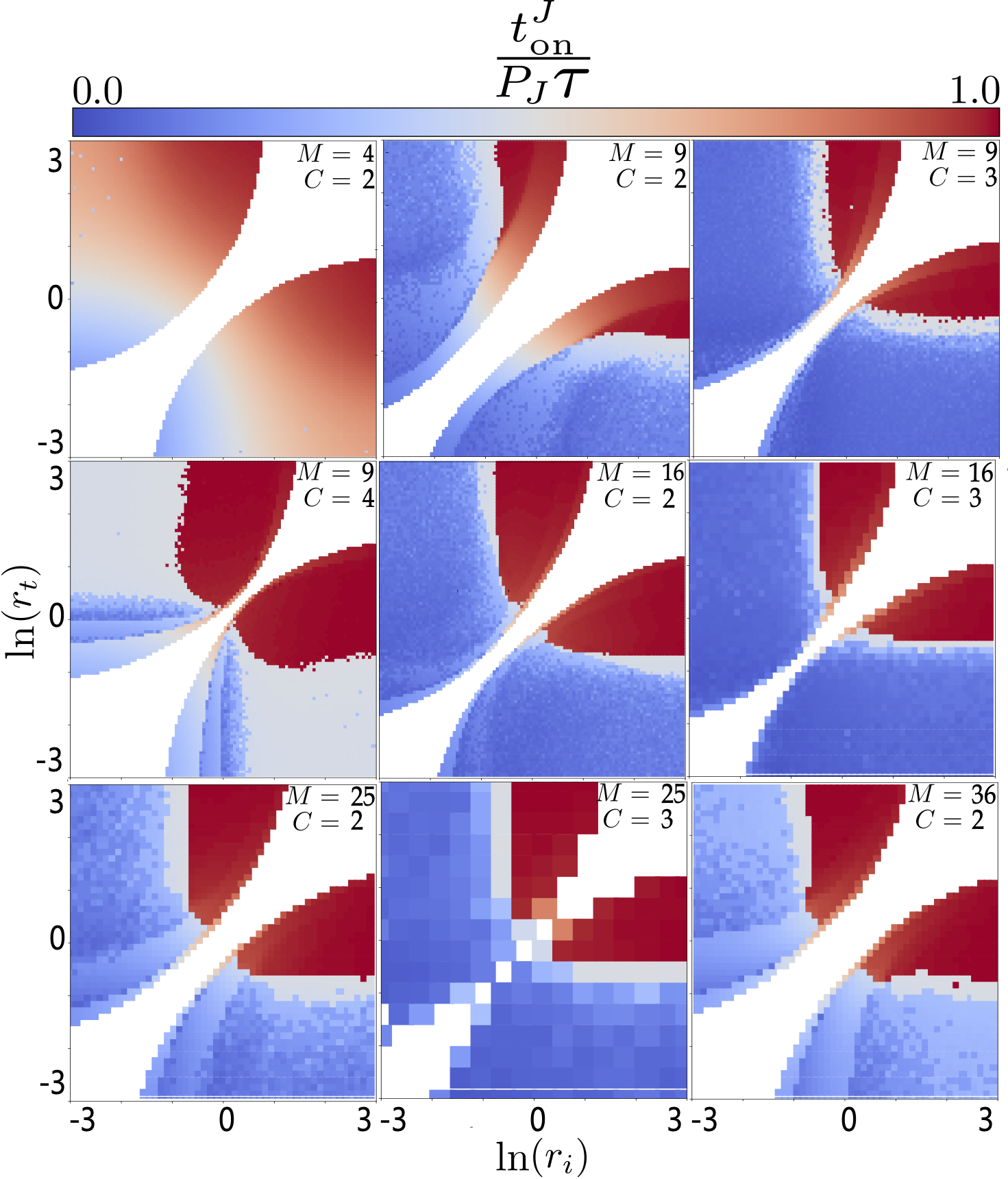}
\includegraphics[width= 0.495\textwidth,height = 0.44\textheight]{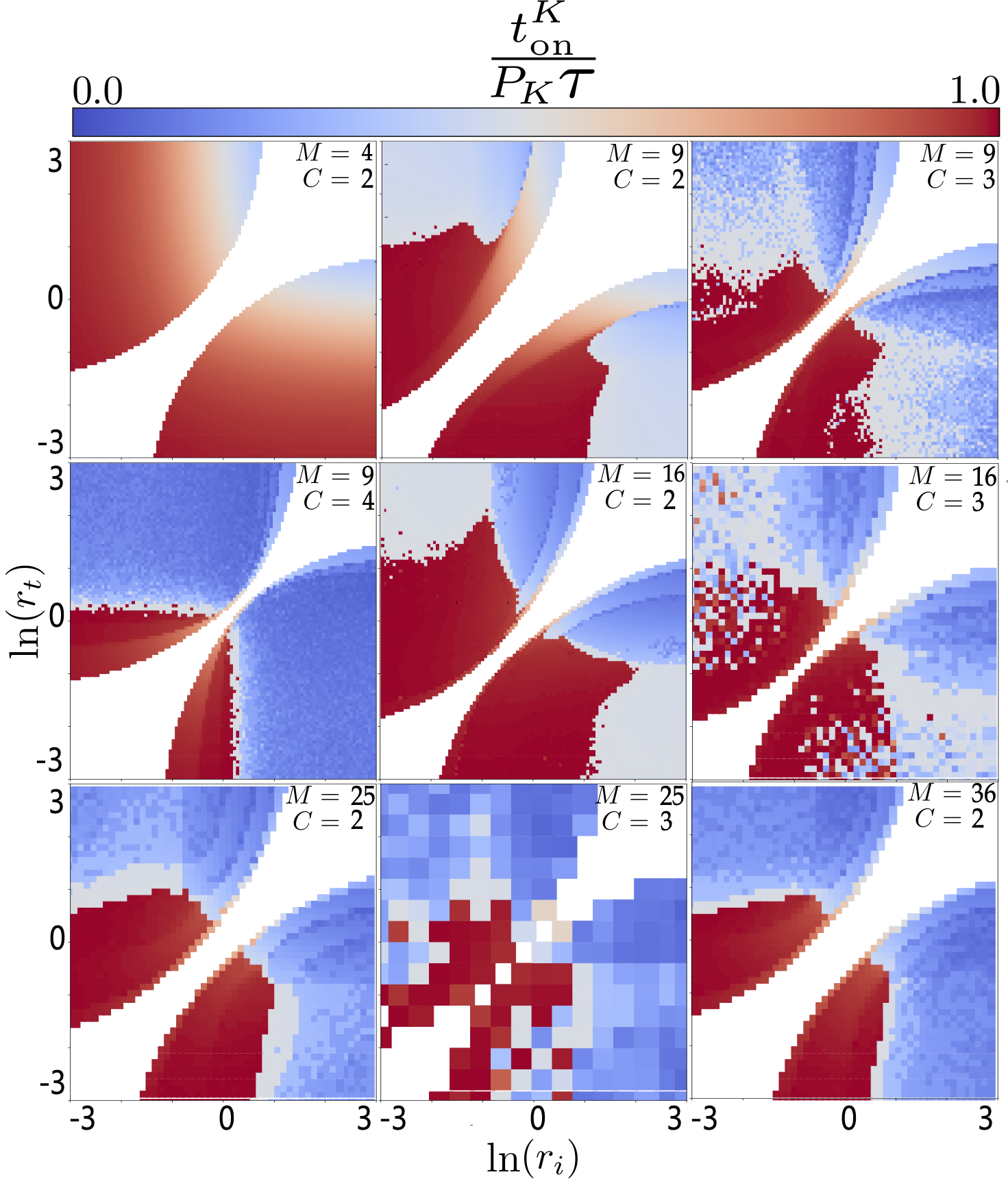}
\caption{The characteristic "on" times for all nine system sizes explored. As the dimension $d =$ $M\choose C$ increases, we decrease the resolution due to computational complexity. White space indicates no data due to the initial and target states being nearly identical}
\label{char_time}
\end{figure*}

\begin{figure}[h!]
\includegraphics[scale=0.45]{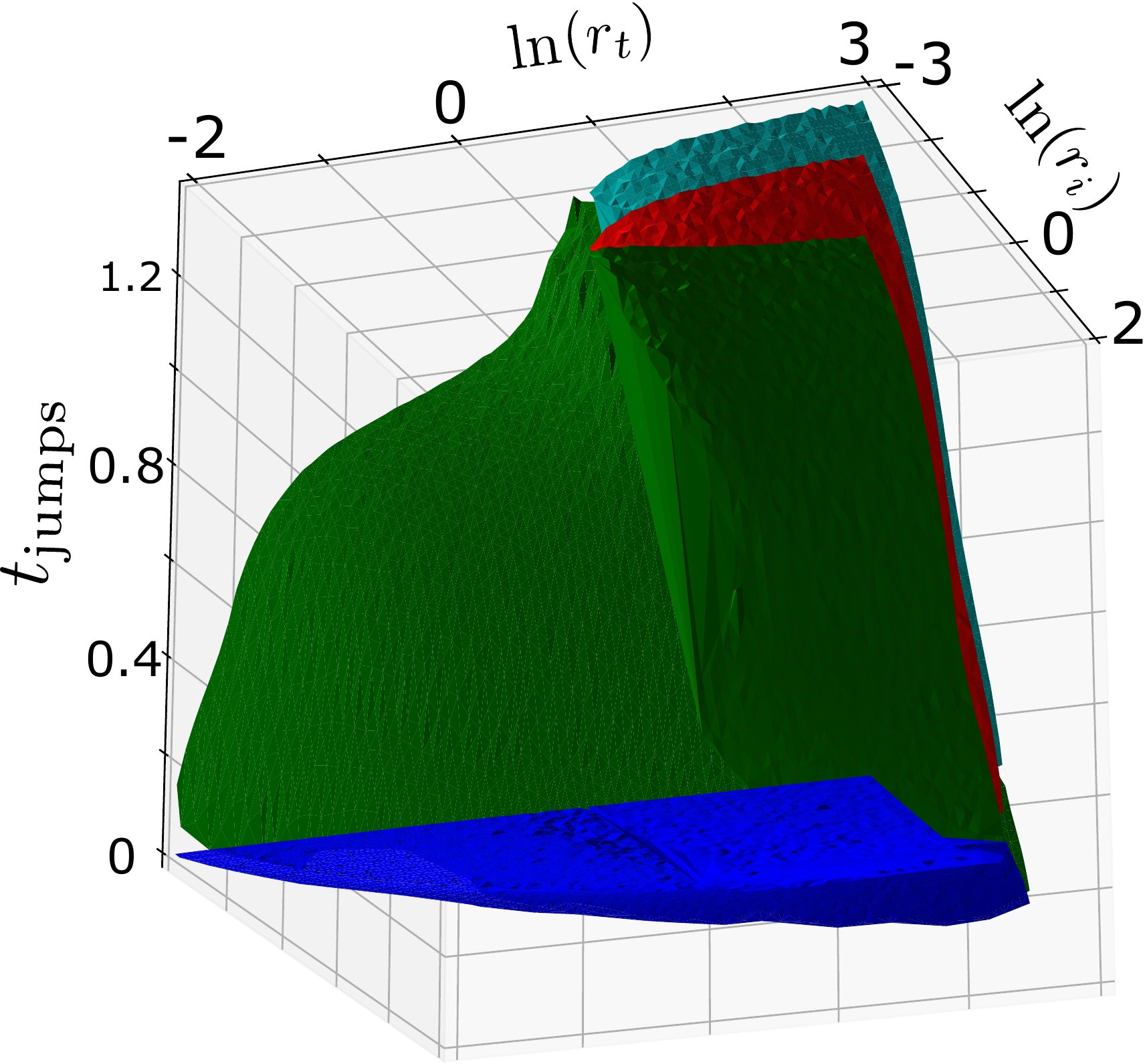}

\vspace{0.2cm}

\includegraphics[scale=0.35]{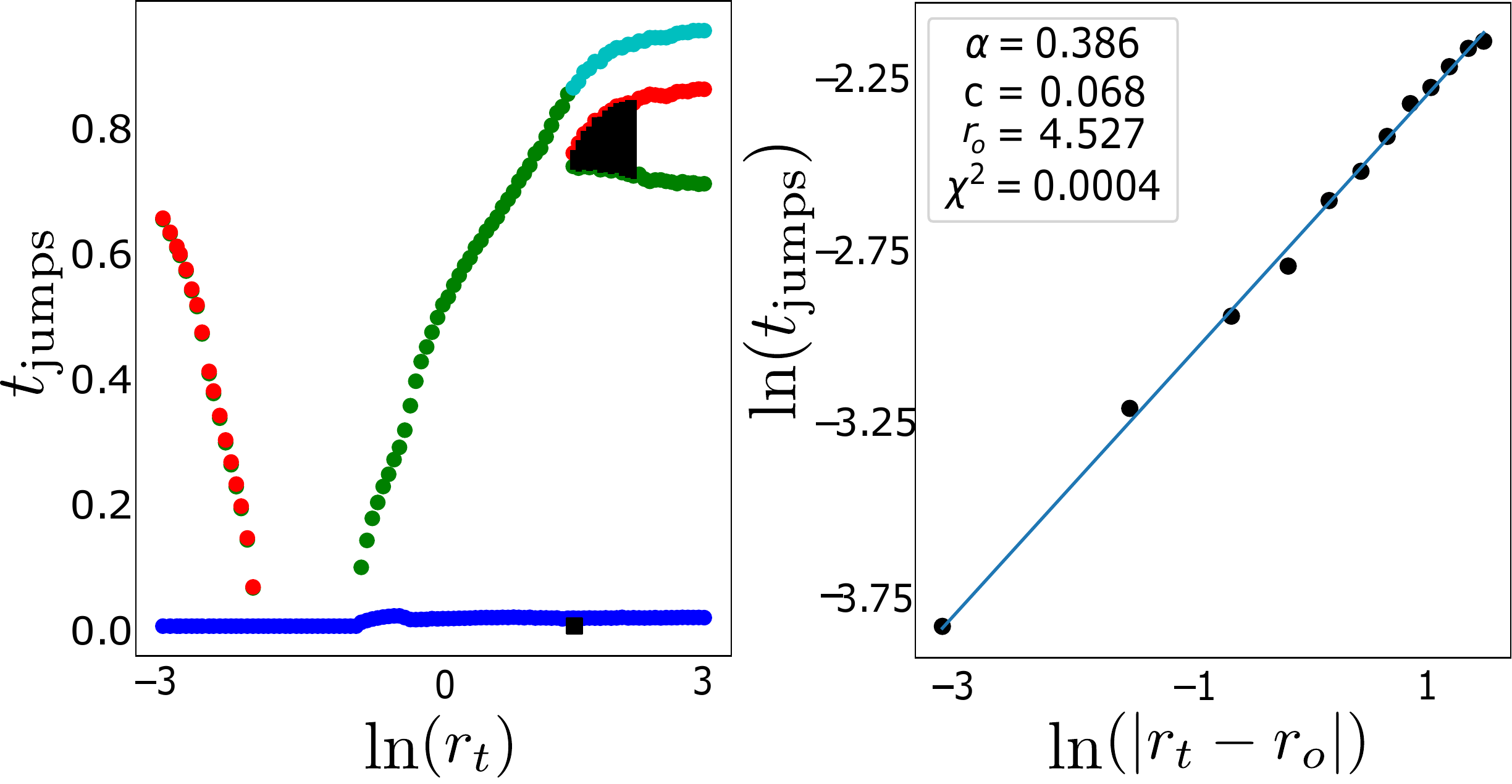}

\caption{{Top:} The optimal protocols in $K$ for $M=9,\ C=2$, restricting the image to $r_i < r_t$. Colors indicate the orders of jumps, with blue first, green, red, and cyan last. The optimal protocol seems to be a continuous function of $r_i, r_t$. A bifurcation opens (red and green surfaces) where $\ln(r_t) = 1$.
{Bottom:} A cross section of the above plot where $\ln(r_i) = -1.4$, along with a power-law fit $t_{\rm pulse} = (r_t - r_0)^{\alpha} + c$ with three fitting parameters, $\alpha$, $r_0$, and $c$.}
\label{bifurcation}
\end{figure}

\section{Topological phase diagram of the optimal protocols}
\label{sec:topological}
Bang-bang protocols are characterized by one integer, namely, the number of pulses in the protocol. This topological property is associated with every protocol in the $(r_i, r_t)$ space. Thus the above space breaks into equivalence classes, each with a fixed number of on pulses. These regions of the $(r_i, r_t)$ space are reminiscent of different topological phases. We thus refer to them as a topological phase diagram. The analogy might appear superficial at this stage. However, the emergence of critical exponents at the transitions between these regions and geometric correlations between protocols within one region suggest a possibly deeper relationship.

Close to the diagonal, we seem to have only one on pulse in both $J$ and $K$ (Fig. \ref{jumps}). As we move away from the diagonal to regions with a smaller overlap and a longer critical total time, we see an increase in the number of pulses. The number of pulses is correlated with the critical time.

The number of pulses changes by 1, going from $P$ to $P+1$, as we cross a phase boundary. Thus the diagram has a layered structure, where phases with $P+1$ pulses appear as islands enclosed by phases with $P$ pulses. This feature can be explained by noticing that the transition mechanism is through a bifurcation. As an example, consider an interval where a control is turned off. At the transition, an infinitesimally narrow square on pulse occurs at some point in this interval. The width of the pulse emerging at the transition grows continuously from zero. Interestingly, there are many similarities in the structure of the phase boundaries in the number of pulses and the overlap $|\langle \psi_{\rm target}|\psi_{\rm initial}\rangle|^2$.

Another pattern emerges in the characteristic time for on pulses, $t_{\rm on}^J/(P_J \tau)$  and $t_{\rm on}^K/(P_K\tau)$. The upper right quadrant with $\ln(r_i), \ln(r_t) > 0$ has a single constant on pulse in $J$ across all system sizes explored, with a similar pattern in $K$ where $\ln(r_i), \ln(r_t) < 0$ shown in Fig. \ref{char_time}. From these times, where $t_{\rm on}^J/(P_J \tau) = 1$, we transition into  $t_{\rm on}^J/(P_J \tau) = 0.5$, which signals a bifurcation opening up, with smooth transitions to $t_{\rm on}^J/(P_J \tau) < 0.5$. The data also suggest that $r_ir_t > 1$ results in $J$-dominant protocols, with $r_ir_t < 1$ resulting in $K$-dominant protocols. {\color{black} We note that the number of pulses (Fig.~\ref{jumps}) and the typical timescale of the each pulse (Fig.~\ref{char_time}) reflect different and complementary
aspects of the protocols. For example, the red region in the $P_J$ plot for $M=9$ and $C=3$ in Fig.~\ref{jumps} indicates many pulses in $J$. However, the complementary panels in Fig. \ref{char_time} indicate that these are short $J$ pulses and the dynamics are actually dominated by fewer but longer $K$ pulses.}

The continuous nature of the bifurcations raises the question of any connection to critical phenomena. Interestingly, the duration of the pulses that appear at the bifurcation transition grows as a power law for all transitions in the phase diagram, as shown, e.g., in Fig. \ref{bifurcation}. A representative three-dimensional plot of the optimal protocols in $K$ for $M =9, C=2$ is shown in Fig. \ref{bifurcation}. Different surfaces indicate the times of jumps in the bang-bang protocols. We see continuous changes in the optimal protocol as a function of $r_i, r_t$.

Searching for universality, we investigated these power laws for many different bifurcations. While generally there is a good critical fit for all bifurcations, we have not been able to find a universal exponent governing the transitions throughout the phase diagram. The exponents may be analogous to other continuously changing critical exponents, e.g., in a Luttinger liquid.

\section{Characterizing Geometric  correlations between bang-bang Protocols}
\label{sec:geometric}
To further scrutinize the analogy between the region of the $(r_i, r_t)$ space with phases, we note that in a ground-state phase diagram, states within a phase have unifying properties. In addition to the topological pulse number above, each protocol has a geometric structure associated with the precise times the control is turned on and off. Is the geometry correlated within each phase?

To capture the geometric similarity, we need to quantify it in terms of a correlation function. We define 
\begin{align*}
    C[a(t), b(t)] \equiv \int_0^1 {1\over 2}\left\{[2a(t)-1][2b(t)-1]+1\right\} dt
\end{align*} 
for normalized protocols $a(t), b(t)$. This function measures the fractional overlap of bang-bang protocols where the values of $a(t), b(t)$ are restricted to 1 or 0 at any given time $t$.  $C[a(t), b(t)] = 1$ implies identical normalized protocols, and $C[a(t), b(t)] = 0$ implies perfectly anticorrelated protocols. We note that perfect anticorrelation is only possible for two protocols with the same number of jumps occurring at the same normalized time.

\begin{figure*}
\includegraphics[width= 0.495\textwidth,height = 0.45\textheight]{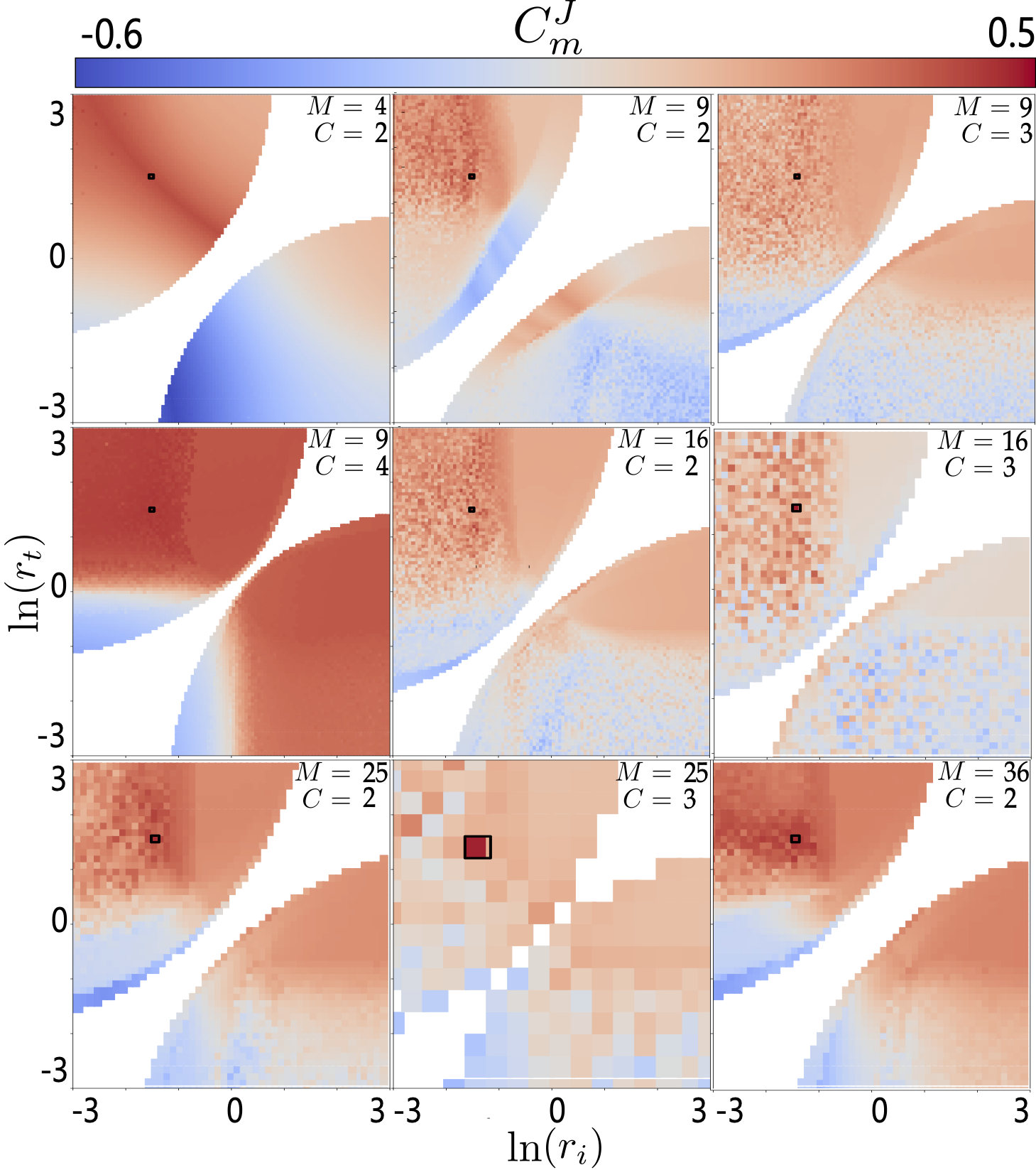}
\includegraphics[width= 0.495\textwidth,height = 0.45\textheight]{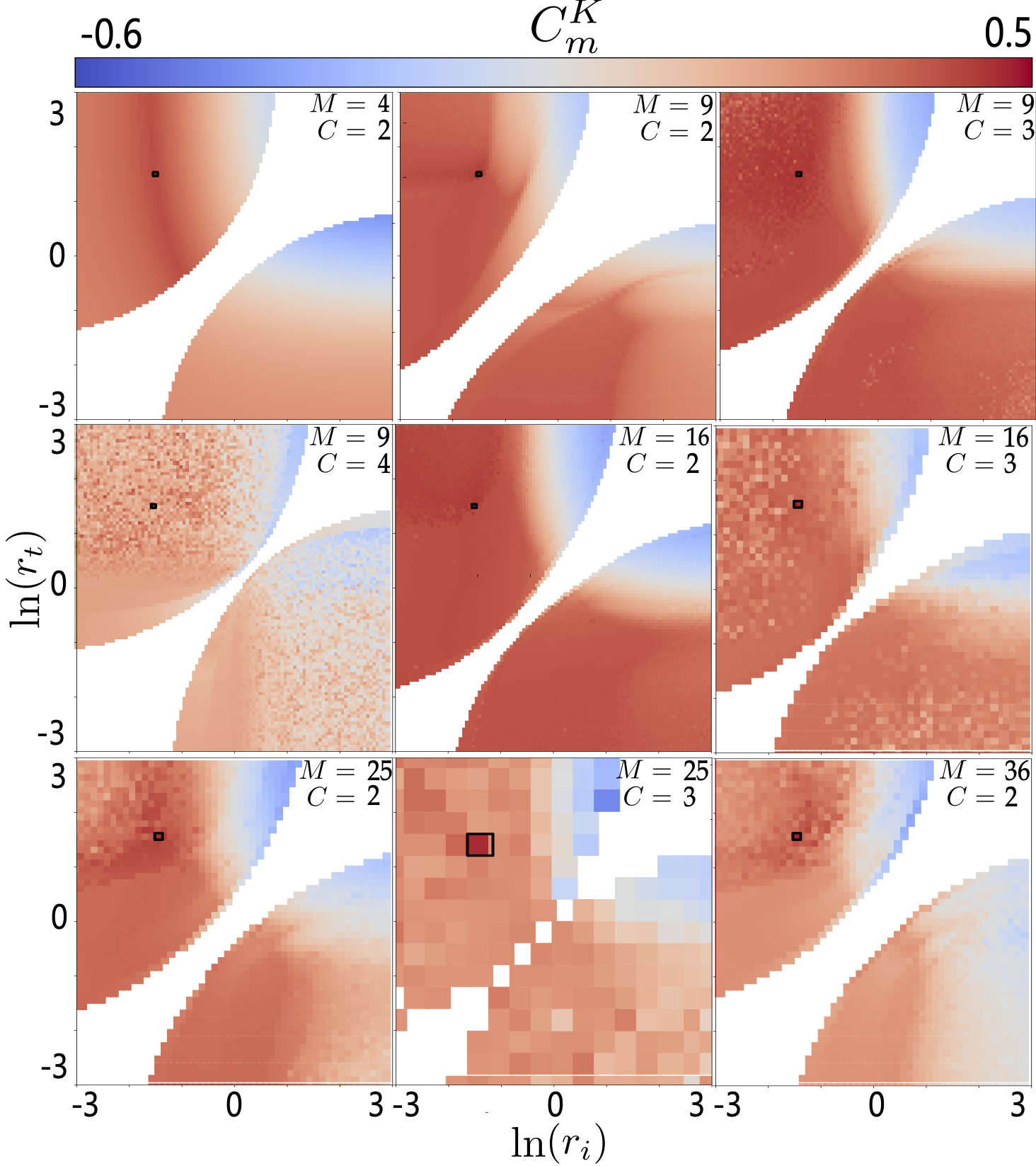}
\includegraphics[width= 0.495\textwidth,height = 0.45\textheight]{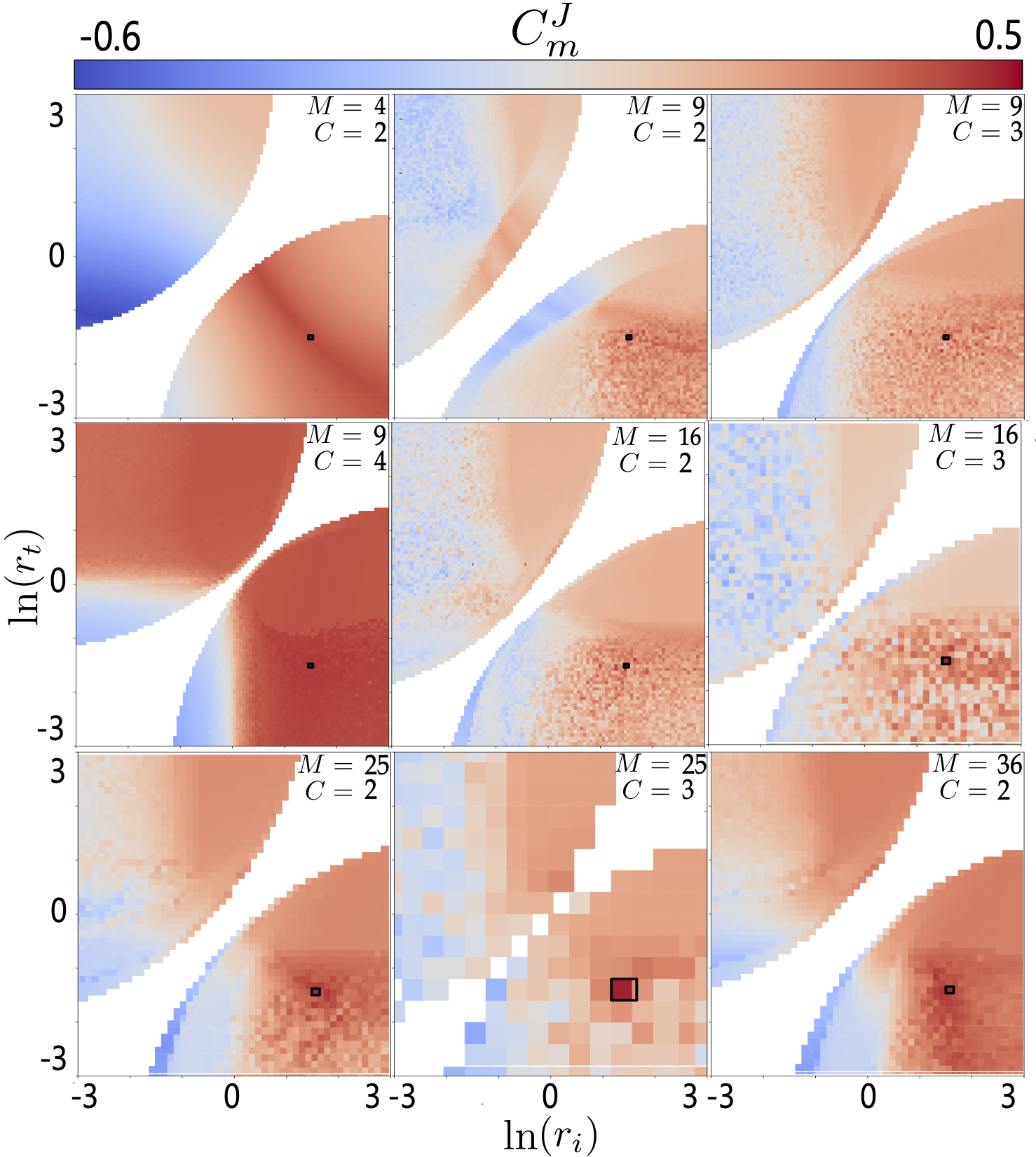}
\includegraphics[width= 0.495\textwidth,height = 0.45\textheight]{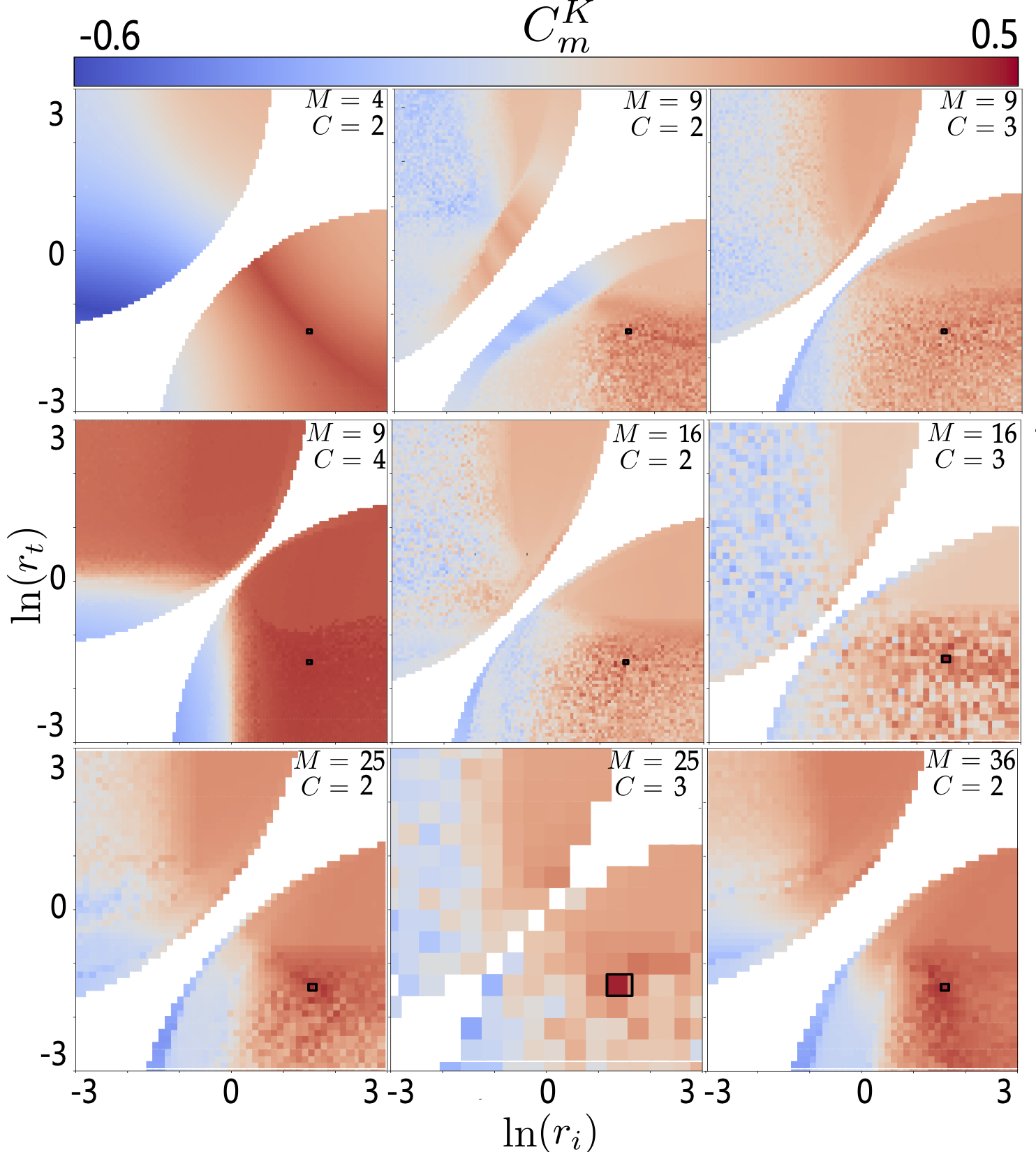}
\caption{The protocol correlations for all nine system sizes. Black squares indicates the protocol which is being compared to. As the dimension $d =$ $M\choose C$ increases, we decrease the resolution due to computational complexity. White space indicates no data due to the initial and target states being nearly identical.}
\label{correlation1}
\end{figure*}

Accounting for the fact that the expected output of $C$ varies based on the number of jumps in $a(t)$ and $b(t)$, we introduce the modified correlation function
\begin{align*}
C_m[a(t), b(t)] \equiv C[a(t), b(t)] - \overline{C[S]}
\end{align*} 
to effectively subtract the background. The above correlation function calculates the difference between the protocol overlap and the average protocol overlap given the total number of jumps, $S$, in both protocols. To calculate $\overline{C[S]}$, we randomly draw $S$ total jumps from the interval $[0,1]$, sort the times of jumps in the protocols, and let $s_i$ correspond to these sorted times. Then the two protocols have the same value on the intervals $[s_{2n}, s_{2n+1}]$. For even $S$, setting $s_{S+1} = 1$, and considering $S!$ possible orderings for these sorted times, $\overline{C[S_{\rm even}]}$ can be calculated as
\begin{align*}
   \overline{C[S_{\rm even}]} = & S! \int_0^{s_{S+1}}  \dots \int_0^{s_2} \sum_{i=1}^{S+1} -1^{i+1}s_i  \ ds_1 \dots ds_K\\
    =& S! \sum_{i=1}^{S+1}(-1)^{i+1} \frac{i}{(S+1)!}
    = \frac{S+2}{2(S+1)}.
\end{align*}

In the case of odd $S$,
\begin{align*}
\overline{C[S_{\rm odd}]} = \frac{1}{S+1}\sum_{i=1}^{S}(-1)^{i+1} i = {1\over 2},
\end{align*} 
where the sum only goes up to $S$ rather than $S+1$ because $[s_S,1]$ is now an anticorrelated region. As a check, we numerically generated $10^6$ random protocols for all $S \leq 10$, and calculated the average $C$, which was in agreement with the expression above.

The behavior of the correlation function is shown in Fig. \ref{correlation1}. We compare all protocols to two different protocols for each system size, which are outlined in black. We find that protocols within one phase exhibit correlations, while anticorrelations emerge across the phase boundaries.

\section{conclusions}
\label{sec:conclusions}

In this paper, we studied patterns in the optimal protocols scanning over a wide range of initial and target ground states of the two-dimensional $XXZ$ model for various system sizes. Identifying patterns and properties of the optimal protocols and characterizing the needed timescales are crucial for efficiently implementing VQA on near-term hybrid quantum devices.

To achieve the above goal in the first stage, where the quantum evolution is simulated on classical computers, we pushed the state of the art substantially by several algorithmic inventions and optimizations. These improvements enabled us to tackle an exceedingly challenging problem of finding globally optimal protocols for nonequilibrium state transformation in a truly many-body setup with large Hilbert spaces for a two-dimensional interacting system. 

The complexity of VQA ultimately relies on the critical time needed for transforming a quantum state to the target with an optimal protocol. The longer this time, the more challenging it gets to find the optimal protocol even with a quantum device that can generate the time evolution (instead of calculating it as in this paper). 
{\color{black}We found that for the $XXZ$ model on the square lattice, the total time does not necessarily increase with the Hilbert space dimension. Although this counterintuitive finding is specific to the case studied here, it is highly encouraging for future applications of VQA. }

 We also find that the wave -unction overlap seems to be the key determinant of the critical time. The overlap is a measure of distance in the Hilbert space, imposing a fundamental speed limit even if we could generate a direct rotation in the Hilbert space. Such direct rotation typically requires nonlocal generators. The fact that the optimal time for dynamics generated by a physically relevant local device Hamiltonian also correlated with the overlap is a promising indicator of the potential of VQA. { \color{black} Similarly, we found this result for the specific $XXZ$ model. Thus it remains an open question whether the correlation between wave-function overlap and the critical preparation time is a generic property of many-body interacting systems, which calls for future investigations on other models. Fermionic and magnetically frustrated systems are of particular interest in this regard. Nevertheless, the $XXZ$ model is nonintegrable and does not map to any noninteracting models. It therefore appears that our model-specific findings might apply to a broad class of interacting systems.  }

We introduced the notion of a phase diagram for the optimal protocols in the space of initial and target states. Since each optimal bang-bang protocol is characterized by an integer number of pulses, the space breaks into regions of the same pulse number. These topological phases are separated by continuous bifurcation transitions and exhibit a layered structure. The number of pulses goes up upon increasing critical preparation time.

We also introduced a correlation function to capture the geometric similarities of bang-bang protocols and found that the protocols within a phase are geometrically correlated {\color{black}for the $XXZ$ model. These findings can inform efficient VQA implementation along two directions. First, finding optimal protocols for a particular initial and target state can yield excellent initial guesses for other initial and target states for the same system size. It seems natural that small changes to the initial and target states should correspond to small changes in the optimal protocols connecting them regardless of the model.}

More importantly, the changes across system sizes also exhibit a progression that can provide good initial guesses for the optimal protocols for a slightly larger system or slightly lower or higher filling fraction. Our results for the $XXZ$ model suggest that the challenges of applying VQA to large systems may be mitigated by exploring all smaller systems for a range of initial and target states. The topological and geometric patterns in the optimal protocols may be utilized to construct smart initial \textit{Ans\"atze} for the larger systems. {\color{black}This finding calls for further investigations on larger systems beyond the capabilities of classical computers by using hybrid quantum-classical machines while utilizing many of the improvements to the classical optimization algorithm presented in this paper. Such investigations may be transformative for quantum technology. For example, suppose the slow transformation of the optimal protocols with system size persists to the thermodynamic limit. In that case, a system-size adaptive VQA, where the optimal protocols for each system size construct the initial \textit{Ansatz} for the subsequent system size, would yield a true quantum advantage in simulating many-body interacting systems.}

\acknowledgements
A.R. thanks Pedram Roushan for helpful discussions. We are grateful to the College of Science and Engineering and the Computer Science Department at WWU for providing access to computing clusters. We thank Zach Mcgrew for optimizing the cluster usage, which helped speed up our computations. This work was supported primarily by the National Science Foundation under Award No. DMR-1945395.  A.R. is grateful to the Kavli Institute for Theoretical Physics for hospitality during parts of this project, acknowledging support by
the National Science Foundation under Grant No. PHY-1748958.

\begin{appendices}
\section*{Appendix: Algorithm Implementation and Optimization Details}
For a fixed $\tau$ and lower and upper limits on the number of intervals $N_{\min} = 2^i, N_{\max} = 2^{i+j}$, implementation of the BBMC algorithm is as follows:
\begin{enumerate}
    \itemsep-0.2em 

    \item Diagonalize the Hamiltonian for the three meaningful combinations of $J$ and $K$, and store them in $V_{JK}, D_{JK}$.
    \item For each possible total interval number $N_a$ such that $N_{\min} \leq N_a = 2^a \leq N_{\max}$ for some $a$, generate and save the unitary matrices for each combination of timestep $\triangle t = {\tau}/{N_a}$ and $H_{JK}$ . This results in $3j$ total unitary matrices where $j = \log_2(\frac{N_{\max}}{N_{\min}})$. Restricting the total number of intervals to a power of some fixed integer $b$ allows the optimal protocols for $N_a=b^a$ to be used as an initial protocol for $N = b^{a+1}$.
    \item Start with $N_a = N_{\min}$ and some random initial protocol. Use the standard annealing process outlined by the BFMC, where a random interval selection now switches the protocol's value at that time. 
    \item Double the number of steps and repeat step 3. Use the optimal protocol for the previous step size as an initial protocol in the next step size. Do this until $N_a = N_{\max}$.
    \item Convert the optimal protocol for $N_a = N_{\max}$ into one that specifies the time that jumps occur. With this conversion, run a second simulation that performs a similar annealing process, except that it now randomly selects the time that the jumps occur and makes some change in that time that is proportional to $T$.
    \item Repeat steps  2-5, scaling time, until $\mathcal{D}[\psi(\tau)] \leq \epsilon$.
\end{enumerate}
Too few steps make the evolution coarse and restrict the time that these jumps can occur. Too many steps make the DBMC computationally expensive and make it difficult to find the optimal protocol with so many indices to choose from. We find that $N_{\min} = 4, N_{\max}= 64$ is enough to get us close to the optimal protocol without getting stuck in local minima. Each time the number of steps is increased, the initial protocol for the next DBMC run is the optimal protocol for the previous step size, which reduces the total number of sweeps required.

This adaptive step size also allows for another efficiency boost. For $n \geq 2N_{\min}$, if a given protocol is unchanged for many steps, use the larger timestep exponentiated matrix. We find that most optimal protocols are fixed for many timesteps. When $n = N_{\max}$ we end up doing significantly fewer than $N_{\max}$ matrix-vector multiplications during evolution.
Several optimization techniques are used to increase the speed of the computation as discussed below.

{\it Scaling total time.} We choose a fixed initial time. After the first iteration of the BBMC process, it linearly extrapolates the total time we need to get $\mathcal{D}[\psi(\tau)] \approx 0.2$. After this, it scales total time after each iteration, with the scalar being roughly proportional to the distance to our target.

{\it Adaptive step size.}
Early on in the CBMC processes, especially when near our random initial protocol, it is necessary to make significant changes in the protocol. When near the optimal protocol, small changes are required, as it is unlikely that large changes will lead to improvement. To achieve this, we set a temperature-dependent upper bound $B(T)$ for the allowed change, which starts off as a significant fraction of total time, usually $B(T_0) = 0.8\tau$ and decays at the same rate as the pseudotemperature to less than 2\% of the total time. For a fixed upper bound $B$, we randomly draw a change from $[0, B(T)]$.

{\it Varying total sweeps.}
For total times much shorter than $\tau_{\rm critical}$, 
convergence is relatively easy and requires few sweeps. As we approach the critical time, with $\mathcal{D}[\psi(\tau)]$ approaching 0, convergence becomes more difficult. Furthermore, iterations with fewer variational parameters need significantly fewer sweeps. To account for these issues, we allow the total number of sweeps to be proportional to the total number of intervals or jumps.

{\it Saving each state during evolution.}
During all MC simulations, changing a given interval does not change the state leading up to that interval. Therefore we can save the state at every step in the evolution and only ``continue'' the evolution from the change onward. Since the interval that receives the change is uniformly distributed across all possible steps, this cuts the computation time by a factor of 2.

{\it Penalizing fictitious jumps.}
To save steps during the BFMC and DBMC, we bias our index selection towards points near a jump. It is unlikely that a single interval getting changed in the middle of a plateau is going to get us closer to the target state:  Progress is more likely to be made by slightly shifting the time a jump occurs. To implement this bias, we add a ``reroll'' if an index is selected, which has identical neighbors.  To further prevent wasted iterations, we do not allow changes that result in both $J$ and $K$ being turned off.

\end{appendices}
\bibliography{citations.bib}
\end{document}